\documentclass[12pt]{article}
\usepackage[dvips]{graphicx}
\usepackage{amsfonts,amssymb,amsthm,amstext,amscd}

\newcommand{\beqa}{\begin{eqnarray}}
\newcommand{\eeqa}{\end{eqnarray}}

\def\d{{\rm d}}

\newcommand{\be}{\begin{equation}}
\newcommand{\ee}{\end{equation}}
\newcommand{\beq}{\begin{equation}}
\newcommand{\eeq}{\end{equation}}
\newcommand{\bea}{\begin{eqnarray}}
\newcommand{\eea}{\end{eqnarray}}

\newcommand{\rc}{\nonumber\\}
\newcommand{\bear}{\begin{eqnarray}}
\newcommand{\eear}{\end{eqnarray}}
\begin{document}
\baselineskip=15.5pt
\pagestyle{plain}
\setcounter{page}{1}

\def\r{\rho}   
\def\CC{{\mathchoice
{\rm C\mkern-8mu\vrule height1.45ex depth-.05ex
width.05em\mkern9mu\kern-.05em}
{\rm C\mkern-8mu\vrule height1.45ex depth-.05ex
width.05em\mkern9mu\kern-.05em}
{\rm C\mkern-8mu\vrule height1ex depth-.07ex
width.035em\mkern9mu\kern-.035em}
{\rm C\mkern-8mu\vrule height.65ex depth-.1ex
width.025em\mkern8mu\kern-.025em}}}


\def\cst{c_T}
\def\csx{c_X}
\def\csr{c_R}
\def\css{c_S}
\def\csf{c_F}
\def\csu{c_U}

\def\ht{{\hat{t}}}
\def\hxm{{\hat{x}^\mu}}
\def\hxn{{\hat{x}^\nu}}
\def\hr{{\hat{r}}}
\def\ha{{\hat{a}}}
\def\hb{{\hat{b}}}
\def\hc{{\hat{c}}}
\def\hd{{\hat{d}}}
\def\htau{{\hat{\tau}}}
\def\hi{{\hat{x}^i}}
\def\hj{{\hat{j}}}
\def\hK{{\hat{K}}}
\def\hL{{\hat{L}}}
\def\hM{{\hat{M}}}
\def\hN{{\hat{N}}}
\def\d{\partial}
\def\med{\frac{1}{2}}

\newfont{\namefont}{cmr10}
\newfont{\addfont}{cmti7 scaled 1440}
\newfont{\boldmathfont}{cmbx10}
\newfont{\headfontb}{cmbx10 scaled 1728}
\renewcommand{\theequation}{{\rm\thesection.\arabic{equation}}}
\font\cmss=cmss10 \font\cmsss=cmss10 at 7pt
\par\hfill KUL-TF-09/17

\begin{center}
{\LARGE{\bf D3-D7 Quark-Gluon Plasmas}}
\end{center}
\vskip 10pt
\begin{center}
{\large 
Francesco Bigazzi $^{a}$, Aldo L. Cotrone $^{b}$, Javier Mas $^{c}$, \\
Angel Paredes $^{d}$, Alfonso V. Ramallo $^{c}$, Javier Tarr\'\i o  $^{c}$.}
\end{center}
\vskip 10pt
\begin{center}
\textit{$^a$ Physique Th\'eorique et Math\'ematique and International Solvay
Institutes, Universit\'e Libre de Bruxelles; CP 231, B-1050
Bruxelles, Belgium.}\\
\textit{$^b$  Institute for theoretical physics, K.U. Leuven;
Celestijnenlaan 200D, B-3001 Leuven,
Belgium.}\\
\textit{$^c$ Departamento de  F\'\i sica de Part\'\i culas, Universidade de Santiago de Compostela and Instituto Galego de 
F\'\i sica de Altas Enerx\'\i as (IGFAE); E-15782, Santiago de Compostela, Spain.}\\
\textit{$^d$ Departament de F\'\i sica Fonamental and ICCUB Institut de
Ci\`encies del Cosmos, Universitat de Barcelona, Mart\'\i\ i Franqu\`es, 1,
E-08028, Barcelona, Spain.
}\\
{\small fbigazzi@ulb.ac.be, cotrone@itf.fys.kuleuven.be, javier.mas@usc.es, aparedes@ffn.ub.es, alfonso@fpaxp1.usc.es, tarrio@fpaxp1.usc.es}
\end{center}

\vspace{15pt}

\begin{center}
\textbf{Abstract}
\end{center}

\vspace{4pt}{\small \noindent 
We present the string dual to finite temperature $SU(N_c)$ ${\cal N}=4$ SYM coupled to massless fundamental matter introduced by $N_f$ D7 branes,
with Abelian flavor symmetry. The analytic solution includes the backreaction of the flavors up to second order in the
parameter that weighs the internal flavor loops,
$\epsilon_h=(\lambda_h N_f)/(8\pi^2 N_c)$, $\lambda_h$ 
being  the 't Hooft coupling at the temperature of the dual Quark-Gluon Plasma. We study the thermodynamics of the system and its departure from conformality, which is a second order effect. We then analyze the energy loss of partons moving through the plasma, finding that the fundamental degrees of freedom enhance the jet quenching. The whole setup is generalized to D3-D7 systems with D3-branes placed at the tip of a generic singular Calabi-Yau cone over a five-dimensional Sasaki-Einstein manifold. We finally provide the equations for the inclusion of massive flavors in the ${\cal N}=4$ SYM plasma.

}
\vfill

\newpage


\section{Introduction}
\setcounter{equation}{0}

The ``fireballs'' experimentally produced at RHIC and the ones that will be produced at the LHC contain a non-zero fraction of degrees of freedom in the fundamental representation \cite{arsene}: they are truly Quark-Gluon Plasmas rather than just Gluon Plasmas.
While holographic methods provide interesting tools to analyze these systems, they have mainly  been  concerned with plasmas without flavors in fundamental representations, or have treated the latter in the quenched approximation.\footnote{Exceptions to this statement can be found in \cite{Casero:2005se,Casero:2006pt,noncrit}.} 
The simplest and best studied example is the plasma of ${\cal N}=4$ SYM which, unexpectedly, has proven to be not very different from the QCD one.

This paper is concerned with the study of flavor effects in the planar,
strongly coupled, $SU(N_c)$ ${\cal N}=4$ SYM plasma beyond the quenched approximation.
A solution is presented in section \ref{sec: massless}, dual to ${\cal N}=4\ $ SYM coupled to $N_f\gg 1$ massless flavors.
The latter are introduced by means of D7-branes in the gravity background.
The branes are homogeneously smeared over the transverse directions of the geometry and so the flavor symmetry group is a product of Abelian factors \cite{Bigazzi:2005md,Casero:2006pt}.\footnote{D3-D7 solutions at zero temperature,
where the D7's are localized rather than smeared have been discussed in \cite{localizedD3D7}.} 
The construction is immediately extended to any D3-D7 setup with $N_c$ D3-branes at the tip of Calabi-Yau cones over Sasaki-Einstein manifolds. In all the cases the zero temperature gauge theories are taken to preserve ${\cal N}=1$ supersymmetry in 4d.

The solution can be given in closed analytic form up to second order in 
$\epsilon_h$, which is essentially $\lambda_h N_f/ N_c$ (precise definitions will be given below),
where $\lambda_h$ is the 't Hooft coupling at the temperature of the plasma.
The validity of the approach requires $\epsilon_h$ to be small (see section \ref{sec: hierarchy}), which is a common feature of D3-D7 systems. Actually
in the particular example that we work out in the present paper, realistic inputs yield values of $\epsilon_h \sim 0.24$ (see section 4.1.1 below). 
Therefore, keeping ${\cal O}(\epsilon_h^2)$ is already accurate up to the level of a few percent, the third and higher order corrections to the solution being tiny.
On the other hand, it is important to keep the $\epsilon_h^2$ terms since, as we will see, this is the order at which conformality breaking (at the quantum level) affects the thermodynamical observables.

The gauge theories we focus on, in fact, are conformal in the unflavored case and, once we couple them with dynamical flavors, have a UV Landau pole. The latter is mapped to the blowing up of the dilaton and UV pathologies in the gravity solution at a finite energy
scale \cite{Benini:2006hh,unquenchedmesons}. In order to make meaningful physical statements, we will  take the usual point of view of field
theory: if the IR scale in which we are interested is far below the Landau pole scale,
it is possible to make well-defined predictions for the IR physics in terms of IR quantities.
An eventual UV completion affects the results at most in powers of $\frac{\Lambda_{IR}}{\Lambda_{UV}}$ ($\Lambda_{IR,UV}$ being the IR and UV energy scales).

The first plasma property we study in section \ref{thermosec} is the thermodynamics.
Up to first order in $\lambda_h N_f/N_c$ our results for the thermodynamic quantities (entropy, energy, free energy, heat capacity, speed of sound) confirm the probe computations in \cite{myers}.
The second order results provide new insights into the plasma. 
In particular, we verify the breaking of conformality in the thermodynamic observables at second order, such as the departure of the square of the speed of sound from the value $1/3$. Since we neglect higher derivative corrections to our gravity solutions, the usual relation $\eta/s=1/4\pi$ holds trivially \cite{Kovtun:2004de}, giving automatically the value of the shear viscosity.

An interesting result is provided by the study of the energy loss of partons in the plasma in section \ref{jetsection}.
Using the standard procedures \cite{liu,herzog1,gubserdrag,herzog2,aredmas} we find that the energy loss is enhanced by the fundamental fields. In particular, the holographic value of the jet quenching parameter is shifted to higher values with respect to the unflavored plasma, approaching the experimental window.

We finally start considering the inclusion of massive flavors, whose treatment is more challenging.
We are able to provide in section \ref{S5massive} the equations of motion for the system, but we leave their study for future work.
A summary of the results is presented in section \ref{sec: discussion} and technical details of calculations are relegated to three appendices.

\section{The D3-D7 plasma solution with massless flavors}
\setcounter{equation}{0}
\label{sec: massless}
In \cite{Benini:2006hh}, the gravity duals to a class of ${\cal N}=1$ 4d quiver 
gauge theories coupled to massless fundamental flavor fields,
 were found. The gauge theories describe the low energy dynamics at the intersection of $N_c$ ``color'' D3-branes and $N_f$ ``flavor'' D7-branes. The dual supergravity solutions account for the combined backreaction of both stacks of D-branes and thus allow an exploration of  the non perturbative dynamics of the corresponding gauge theories beyond the quenched approximation. 

In the setup the D3-branes are placed at the tip of a Calabi-Yau cone over a 5d Sasaki-Einstein manifold $X_5$. In the absence of flavor branes they source a background whose near horizon limit is $AdS_5\times X_5$: the dual gauge theories are superconformal quivers. 
Just to pick two well known examples: when $X_5=S^5$ the CY manifold is just the 6d Euclidean space and the dual field theory is ${\cal N}=4$ SYM; when $X_5=T^{1,1}$ the CY manifold is the singular conifold and the dual theory is the Klebanov-Witten quiver \cite{Klebanov:1998hh}. 
Since this will be used in the following let us remember that the metric of a 5d Sasaki-Einstein manifold can be written as a $U(1)$ fibration over a 4d K\"ahler-Einstein manifold:
\be
ds_{X_5}^2 = ds_{KE}^2 + (d\tau + A_{KE})^2\,\,,
\label{x5metric}
\ee
where $\tau$ is the fiber coordinate and $A_{KE}$ is the connection one-form whose curvature is related to the 
K\"ahler form of the KE base: $dA_{KE}=2J_{KE}$. 
For $X_5=S^5$ the KE base is $CP^2$ and for $X_5=T^{1,1}$ it is $S^2 \times S^2$.

The D7-branes introduce fundamental matter in the dual field theories. They are extended along the radial direction of the background and wrap a submanifold $X_3$ of $X_5$. They are also homogeneously smeared over the transverse space \cite{Bigazzi:2005md,Casero:2006pt}.\footnote{See \cite{tutti,Benini:2006hh,bcnp,fkw1,hnp,fkw2,Bigazzi:2008qq,unquenchedmesons} for other studies of this kind of construction in the zero temperature case.} 
The smeared distribution is taken in such a way that the isometries of the fibered K\"ahler-Einstein
 space are kept unbroken, and allows
to write an ansatz where all the unknown functions just depend on a single radial coordinate. 
The D7-brane embedding is taken
such that, when the temperature vanishes, the whole D3-D7 system preserves ${\cal N}=1$ supersymmetry in 4d.\footnote{
When $X_5=S^5$, a single stack of parallel D7-branes preserves ${\cal N}=2$. However, the smeared
distributions considered in \cite{Benini:2006hh} and in the present paper just preserve ${\cal N}=1$. See appendix \ref{unquenchedkk} for more details on the
D7-brane embeddings for this case.}
As a general feature of D3-D7 setups, the dilaton runs and blows up at a certain radial distance, 
corresponding to a UV Landau pole in the dual gauge theory \cite{Benini:2006hh}.

In the following, we are going to study the finite temperature behavior of these D3-D7 gauge theories. As a first step, we are going to find the non-extremal generalizations of the supergravity solutions found in \cite{Benini:2006hh}.

\subsection{Ansatz and equations of motion}
\label{sec 21}
We look for solutions of type IIB supergravity coupled to $N_f$ D7-brane sources. The action of the system reads:
\be
S=S_{IIB} + S_{fl}\,\,,
\label{fullaction}
\ee
where the active terms of $S_{IIB}$ are (we will work in Einstein frame throughout the paper):
\be
S_{IIB}=\frac{1}{2\kappa_{10}^2}\int d^{10}x \sqrt{-g_{10}} \left[ R-\med \partial_M\Phi \partial^M\Phi -\med e^{2\Phi}F_{(1)}^2 - \frac12
\frac{1}{5!} F_{(5)}^2 \right]\,, \label{actiongrav}
\ee
whereas the action for the D7 flavor branes takes the usual DBI+WZ form:
\be
S_{fl} = -T_7 \sum_{N_f} \left( \int d^8x\, e^\Phi \sqrt{-g_8}\, - \int C_8 \right)\,.
\label{actionflav}
\ee
The gravitational constant and D7-brane tension are, in terms of string parameters:
\be
\frac{1}{2\kappa_{10}^2} = \frac{T_7}{g_s} = \frac{1}{(2\pi)^7g_s^2 \alpha'^4}\,\,.
\label{kappa10}
\ee
The metric ansatz we want to focus on is:
\be
ds_{10}^2 = h^{-\frac12}\left[-b\ dt^2 + d\vec{x}_3^2\right] + h^\frac12
\left[ b\,S^8F^2 d\sigma^2 + S^2 ds_{KE}^2 + F^2 (d\tau + A_{KE})^2
\right]\,\,.
\label{10dmetric}
\ee
The functions $h,b,S,F$ (as well as the dilaton $\Phi$) depend on the radial variable $\sigma$; $b=1$ in the extremal (i.e. zero-temperature) case. In our conventions, $S,F$ have dimensions of length, $b,h$ are dimensionless and
$\sigma$ has dimension length${}^{-4}$.

The non-trivial RR field strengths are:
\be
F_{(5)} = Q_c\,(1\,+\,*)\varepsilon(X_5)\,\,,\quad
F_{(1)} = Q_f \,(d\tau + A_{KE})\,\,,\quad (dF_{(1)}= 2\,Q_f\,J_{KE})\,\,,
\label{f5f1}
\ee
where $\varepsilon(X_5)$ is the volume element of the internal space and $Q_{c}, Q_{f}$ are proportional to
the number of colors and flavors:
\be
N_c = \frac{Q_c\, Vol(X_5)}{(2\pi)^4g_s \,\alpha'^2}\,\,,\qquad
N_f = \frac{4\,Q_f\,Vol(X_5)}{Vol(X_3) g_s}\ .
\label{NcNf}
\ee
The first relation is just the usual quantization condition of the D3-brane charge, while the
derivation of the second is detailed in appendix \ref{Qf}.\footnote{We have defined $N_f$ as the number of flavor branes. The precise way in which they introduce fundamental degrees of freedom and how they couple to adjoints and bifundamentals depends on the particular theory. For instance, in the $X_5 = S^5$ case, 
there are $2N_f$ supermultiplets (in ${\cal N}=1$ language) but the flavor symmetry is just $U(1)^{N_f}$,
due to the coupling between adjoint and fundamental superfields in the superpotential.}
$Vol(X_3)$ is the volume of the submanifold  ($X_3 \subset X_5$) wrapped by any D7-brane.\footnote{Notice that, since the different D7-branes
take different positions in the internal space due to the smearing, the particular submanifold $X_3$ spanned by each
brane is different. However, since the family of relevant $X_3$'s are related by the internal isometries, they
all have the same volume and thus $Vol(X_3)$ is a well defined quantity.}
In the $X_5=S^5$ case,  $Vol(X_3)=2\pi^2$.

The fact that the flavors are massless
is encoded in the independence  of $F_{(1)}$ on $\sigma$. See \cite{Benini:2006hh,fkw1} for discussions
on this issue and section \ref{S5massive} for the case of massive flavors when $X_5=S^5$.

All the functions we need to compute depend on a single coordinate $\sigma$, and 
it is possible
to describe the system in terms of a one-dimensional effective action. By directly inserting
the ansatz in the action (\ref{fullaction}), we find:
\bear
S_{eff}&=&\frac{Vol(X_5)V_{1,3}}{2\kappa_{10}^2}\int d\sigma \left(
-\frac12\frac{(\d_\sigma h)^2}{h^2} +12\frac{(\d_\sigma S)^2}{S^2} + 8 \frac{(\d_\sigma F)(\d_\sigma S)}{F\,S}-
\frac12 (\d_\sigma \Phi)^2+\right.\rc
&+&\left.
\frac{(\d_\sigma b)}{2b}\left( \frac{(\d_\sigma h)}{h}+ 8 \frac{(\d_\sigma S)}{S}+ 2 \frac{(\d_\sigma F)}{F} 
\right)+
\right. 
\rc
&+&\left.
24 b\,F^2\,S^6 - 4 b\,F^4\,S^4-\frac12 Q_c^2 \frac{b}{h^2}  -\frac12 Q_f^2 e^{2\Phi}
b\,S^8\, 
-
4Q_f
e^{\Phi} \,b\, S^6\,F^2\right)\,\,.
\label{effeclagr}
\eear
In (\ref{effeclagr}) $V_{1,3}$  denotes the (infinite) integral over the Minkowski coordinates. 
The second derivatives coming from the Ricci scalar have been integrated by parts and, as is  customary,
only the angular part of $F_{(5)}$ is inserted in the $F_{(5)}^2$ term (otherwise
the $Q_c$ would not enter the effective action since, on-shell, 
$F_{(5)}^2=0$ due to the self-duality condition). Notice also that the WZ term does not enter
(\ref{effeclagr})  because it does not depend on the metric or the dilaton; its effect has been taken
into account via the expression for $F_{(1)}$ (we refer the reader to \cite{Benini:2006hh}
for extensive explanations).
The  term proportional to $Q_f$ comes from the DBI contribution in (\ref{actionflav});
more detailed considerations regarding this last term
are relegated to appendix \ref{Qf}.
The equations of motion stemming from the effective action (\ref{effeclagr}) are:
\bear
\d_\sigma^2(\log b)&=&0\,\,,\rc
\d_\sigma^2(\log h)&=&-Q_c^2 \frac{b}{h^2}\,\,,\rc
\d_\sigma^2(\log S)&=& -2 b F^4 S^4 + 6 b F^2 S^6 - Q_f \,e^\Phi b\,F^2\,S^6\,\,,
\rc
\d_\sigma^2(\log F)&=& 4b\,F^4 S^4 - \frac{Q_f^2}{2}e^{2\Phi}b\,S^8\,\,,
\rc
\d_\sigma^2\Phi&=& Q_f^2\,e^{2\Phi}\,b\,S^8 +
4 Q_f\,b\, e^\Phi S^6 F^2\,\,.
\label{bhSeqsmassless}
\eear
It is straightforward to check that these equations solve the full set of Einstein equations
provided the following ``zero-energy'' constraint is also satisfied:
\bear
0&=&-\frac12\frac{(\d_\sigma h)^2}{h^2} +12\frac{(\d_\sigma S)^2}{S^2} + 8 \frac{(\d_\sigma F)(\d_\sigma S)}{F\,S}-
\frac12 (\d_\sigma \Phi)^2+
\rc &+&
\frac{(\d_\sigma b)}{2b}\left( \frac{(\d_\sigma h)}{h}+ 8 \frac{(\d_\sigma S)}{S}+ 2 \frac{(\d_\sigma F)}{F} 
\right)+
\rc
&-&
24 b\,F^2\,S^6 + 4 b\,F^4\,S^4+\frac12 Q_c^2 \frac{b}{h^2}  +\frac12 Q_f^2 e^{2\Phi}
b\,S^8\, 
+
4Q_f
e^{\Phi} \,b\, S^6\,F^2\,\,.
\label{constraintmassless}
\eear
This constraint can be thought of as the $\sigma\sigma$ component of the Einstein equations or, alternatively,
as the Gauss law from the gauge fixing of $g_{\sigma\sigma}$ in the ansatz 
(\ref{10dmetric}). By differentiating
(\ref{constraintmassless}) and using (\ref{bhSeqsmassless}), one finds an identity,
which shows that the system is not overdetermined.
The equations are valid for any K\"ahler-Einstein base space since only the properties 
$R_{ab}^{KE}=6g_{ab}^{KE}$ and $dA_{KE}=2J_{KE}$ are needed when checking the Einstein equations.

\subsection{The supersymmetric solution: second order expansion}

The supersymmetric (zero temperature) solution of (\ref{bhSeqsmassless}), (\ref{constraintmassless}) was found in
\cite{Benini:2006hh}. It corresponds to having $b=1$ and it is dual to ${\cal N}=1$ quiver gauge theories with flavors.
In the present notation, the BPS first
order equations read:
\bear
\d_\sigma h&=-Q_c\,\,,\qquad\qquad\
\d_\sigma S&= S^3 F^2\,\,,\rc
\d_\sigma \Phi&= Q_f\,S^4 e^\Phi \,\,,\qquad\quad
\d_\sigma F&= S^4 F \left( 3-2\frac{F^2}{S^2} - \frac{Q_f}{2} e^\Phi \right)\,\,.
\label{masslessBPS}
\eear
It can readily be shown that these equations, together with $b=1$, are sufficient conditions for 
(\ref{bhSeqsmassless}), (\ref{constraintmassless}) to hold.
In a different, dimensionless, radial coordinate $d\rho= S^4 d\sigma$, the equations for $S,F,\Phi$ can be explicitly integrated, giving  \cite{Benini:2006hh}:
\bear
S&=&\alpha'^{\frac12}\, e^\rho\,\left(1+\epsilon_* (\frac16 +\rho_*-\rho)\right)^\frac16\,\,,\rc
F&=&\alpha'^{\frac12}\, e^\rho\,\left(1+\epsilon_* (\rho_*-\rho)\right)^\frac12
\left(1+\epsilon_* (\frac16 +\rho_*-\rho)\right)^{-\frac13}\,\,,\rc
\Phi&=& \Phi_* -\log(1+\epsilon_*\, (\rho_*-\rho))\,\,,\rc
\frac{dh}{d\rho} &=& -Q_c \,\alpha'^{-2}\, e^{-4\rho} \left(1+\epsilon_* (\frac16 +\rho_*-\rho)\right)^{-\frac23}\,\,,
\label{susysol}
\eear
where for later convenience a scale $\rho_*$ has been introduced and $\Phi_*$ is the value
of the dilaton at that scale. 
We have also inserted powers of $\alpha'$ (which enters as an integration constant of
(\ref{masslessBPS})) in order to give appropriate dimensions.
Notice that the solution is defined for $\rho< \rho_{LP}$ where $\rho_{LP}=\rho_* +\epsilon_*^{-1}$ is the point at which the dilaton blows up. We have chosen to keep the differential equation for $h$, even if it can be solved
in terms of incomplete gamma-functions.
The parameter $\epsilon_*=Q_f\,e^{\Phi_*}$ has been introduced.
It has to be small for the solution to be valid in a large energy range (see
section \ref{sec: hierarchy}) and it will be used as an expansion parameter.
Defining $\lambda_*$ as the 't Hooft coupling\footnote{
\label{lambdafoot}
For the (flavored)
${\cal N}=4$ $SU(N_c)$ theory, the gauge coupling is
$g_{YM}^2=4\pi\,g_s e^\Phi$ (note the choice for the numerical prefactor, which sometimes is taken to be $2\pi$), and thus $\lambda_*=4\pi\,g_s N_c e^{\Phi_*}$. For quiver theories that correspond to different $X_5$ geometries, the gauge groups are of the form $SU(N_c)^n$. Let us generalize a relation from the orbifold constructions
$\sum_i^n 4\pi g_{YM,i}^{-2}= (g_s e^\Phi)^{-1}$ \cite{Klebanov:1998hh,Lawrence:1998ja}, 
and consider all the gauge couplings
$g_{YM,i}$ to be equal. Then $4\pi\,g_s N_c e^{\Phi}$, strictly speaking, gives the 't Hooft coupling at each node of the quiver, divided by $n$. With an abuse of language we will simply refer to it as the 't Hooft coupling.} at the $\rho_*$ scale,
$\epsilon_*$ can be expressed in terms of physical quantities by using (\ref{NcNf}) as:
\be
\epsilon_* = \frac{Vol(X_3)}{16\pi\,Vol(X_5)}\lambda_*  \frac{N_f}{N_c}\,\,,
\label{epsstar}
\ee
and, in particular, $\epsilon_{*\,(X_5=S^5)} = \frac{1}{8\pi^2}\lambda_*  \frac{N_f}{N_c}$.

We have set to 0 the integration constant $c_1$ of \cite{Benini:2006hh} for the sake of IR regularity.
Even if for $c_1=0$, this backreacted geometry still presents an IR singularity (much milder than in the 
$c_1\neq 0$ cases), it is useful to think of the $c_1=0$ solution as the massless limit of a family of
IR regular solutions where the IR singularity is removed by non-zero quark masses
\cite{fkw1}. When we go to finite temperature, the singularity will be hidden behind an event horizon.

In comparing to the unflavored and to the finite temperature solutions, it will be useful to employ an $r$ coordinate
which we define by requiring that $h$ takes the simple and familiar form:
\be
h=\frac{R^4}{r^4}\,\,,\qquad\qquad R^4\equiv \frac14 Q_c=\frac14 N_c \frac{(2\pi)^4 g_s \alpha'^2}{Vol(X_5)}\,\,.
\label{simpleh}
\ee
We can expand $\frac{dh}{d\rho}$ from (\ref{susysol})
and integrate order by order in $\epsilon_*$. Since  (\ref{simpleh}) gives
$h$ explicitly in terms of $r$, this yields an expression for $r(\rho)$. Let us fix the 
additive integration
constant in $h$ such that
 $r(\rho_*)\equiv r_* = \sqrt{\alpha'} e^{\rho_*}$.\footnote{This is different from the choice of integration constant
adopted in \cite{unquenchedmesons}, which required $h$ to vanish at the point where the dilaton
diverges. However, the difference between both integration constants is a quantity suppressed as
$e^{-\frac{4}{\epsilon_*}}$
and its influence on the IR physics is of this order
and therefore negligible with respect to the terms kept in the 
Taylor expansion.} Then:
\bear
r&=&\alpha'^{\frac12} e^\rho\Big[1+\frac{\epsilon_*}{72} \Big(
e^{4\rho-4\rho_*}-1 +12(\rho_*-\rho)\Big)
+\frac{5\epsilon_*^2}{10368}\Bigl(e^{8\rho-8\rho_*} + 6 e^{4\rho-4\rho_*}(3+4(\rho_*-\rho))\rc
&&- (19 -24 (\rho_*-\rho) +144 (\rho_*-\rho)^2) \Bigr) + O(\epsilon_*^3)
\Big]\,\,.
\eear
We can obtain now $F(r),S(r), \Phi(r)$ as expansions up to second order:
\bear
F_0&=&r\Big[1-\frac{\epsilon_*}{24}(1+\frac13 \frac{r^4}{r_*^4})+
\frac{\epsilon_*^2}{1152}\left(17-\frac{94}{9}\frac{r^4}{r_*^4}+\frac59\frac{r^8}{r_*^8}
-48 \log(\frac{r}{r_*})\right)+ O(\epsilon_*^3)\Big]\,\,,\rc
S_0&=&r\Big[1+\frac{\epsilon_*}{24}(1-\frac13 \frac{r^4}{r_*^4})+
\frac{\epsilon_*^2}{1152}\left(9-\frac{106}{9}\frac{r^4}{r_*^4}+\frac{5}{9}\frac{r^8}{r_*^8}
+48 \log(\frac{r}{r_*})\right)+ O(\epsilon_*^3)\Big]\,\,,\rc
\Phi_0&=&\Phi_*+ \epsilon_* \log\frac{r}{r_*} + \frac{\epsilon_*^2}{72}\left(1-\frac{r^4}{r_*^4}
+12 \log\frac{r}{r_*} + 36 \log^2\frac{r}{r_*}\right)+ O(\epsilon_*^3)\,\,,
\label{susymasslesssol}
\eear
where the subscript $0$ means that the solutions are dual to the D3-D7 theories at $T=0$.

\subsection{The non-extremal solutions}
\label{sec:finiteT}

We now look for non-extremal solutions of (\ref{bhSeqsmassless}), (\ref{constraintmassless}), which would provide for a dual description to the finite temperature regime of our D3-D7 gauge theories.
Such solutions are required to  
be regular at the horizon and to
tend to the supersymmetric ($T=0$) ones at energy scales much higher than the temperature.
Concretely, we will require that the geometries coincide with the $T=0$ solutions in the extremal limit and that $F,S,\Phi$ coincide with those in
(\ref{susymasslesssol}) when evaluated at $r=r_*$. This uniquely fixes the order by order expansion
of the non-supersymmetric solution.

The equations for $h$ and $b$ in (\ref{bhSeqsmassless}) are decoupled from the rest and are solved, in
terms of an integration constant $r_h$, by:
\be
b=e^{4r_h^4 \,\sigma}
\,\,,\qquad\qquad
h=\frac{Q_c}{4r_h^4}(1-e^{4r_h^4 \,\sigma})\,\,.
\ee
where $\sigma\in (-\infty,0)$.
We define the $r$ coordinate such that the expression for $h$
(\ref{simpleh}) still holds:
\be
e^{4r_h^4\,\sigma}=1-\frac{r_h^4}{r^4}\,\,.
\label{rcoorddef}
\ee
The extremal limit corresponds to sending the
horizon radius $r_h$ to zero. The metric reads:
\be
ds_{10}^2 = -\frac{r^2}{R^2} (1-\frac{r_h^4}{r^4}) dt^2 + 
\frac{r^2}{R^2} d\vec{x}_3^2 + \frac{R^2\tilde S^8 \tilde F^2}{r^2}  \frac{dr^2}{(1-\frac{r_h^4}{r^4})}
+ R^2 \tilde S^2 ds_{KE}^2 + R^2 \tilde F^2 (d\tau + A_{KE})^2\,\,,
\label{deformedads5bh}
\ee
where:
\be
\tilde S \equiv \frac{S}{r}\,\,,\qquad\qquad\qquad
\tilde F \equiv \frac{F}{r}\,\,.
\ee
We still have to solve for $\tilde F, \tilde S, \Phi$. A straightforward computation from
(\ref{bhSeqsmassless}), (\ref{constraintmassless})
leads to the differential equations for $\tilde F, \tilde S, \Phi$, in terms of the coordinate $r$.
It is easy to check that the $AdS_5$ black hole solution $\tilde F=\tilde S=1$, $\Phi=const$ is 
recovered in the flavorless limit $\epsilon_* =0$. Its deformation,
expanded up to second order in $\epsilon_*$ reads:\footnote{This procedure is analogous to the one adopted in \cite{thermoks}.}
\bear
\tilde F&=&1-\frac{\epsilon_*}{24}(1+ \frac{2r^4-r_h^4}{6r_*^4-3r_h^4})+
\frac{\epsilon_*^2}{1152}\left(17-\frac{94}{9}\frac{2r^4-r_h^4}{2r_*^4-r_h^4}+
\frac59\frac{(2r^4-r_h^4)^2}{(2r_*^4-r_h^4)^2}+\right.\rc
&&\left.-\frac89 \frac{r_h^8 (r_*^4-r^4)}{(2r_*^4-r_h^4)^3}
-48 \log(\frac{r}{r_*})\right)+ O(\epsilon_*^3)\,\,,\rc
\tilde S&=&1+\frac{\epsilon_*}{24}(1- \frac{2r^4-r_h^4}{6r_*^4-3r_h^4})+
\frac{\epsilon_*^2}{1152}\left(9-\frac{106}{9}\frac{2r^4-r_h^4}{2r_*^4-r_h^4}+
\frac59\frac{(2r^4-r_h^4)^2}{(2r_*^4-r_h^4)^2}+\right.\rc
&&\left.-\frac89 \frac{r_h^8 (r_*^4-r^4)}{(2r_*^4-r_h^4)^3}
+48 \log(\frac{r}{r_*})\right)+ O(\epsilon_*^3)\,\,,\rc
\Phi&=& \Phi_*+\epsilon_* \log\frac{r}{r_*} + \frac{\epsilon_*^2}{72}\left(1-\frac{2r^4-r_h^4}{2r_*^4-r_h^4}
+12 \log\frac{r}{r_*} + 36 \log^2\frac{r}{r_*}+\right.\rc
&&\left.
+\frac92   
\left(Li_2(1-\frac{r_h^4}{r^4})-Li_2(1-\frac{r_h^4}{r_*^4})\right)
\right)+ O(\epsilon_*^3)\,\,,
\label{finiteTsol}
\eear
where $Li_2(u)\equiv \sum_{n=1}^\infty \frac{u^n}{n^2}$ is a polylogarithmic function.
The expressions in (\ref{deformedads5bh}), (\ref{finiteTsol}) are the central
results of this paper. Together with (\ref{f5f1}) they provide the full perturbative solution. We are now going to discuss their regime  of validity and, in the following sections, their
implications for the physics of the dual flavored plasmas.

\subsection{Hierarchy of scales and regime of validity}
\label{sec: hierarchy}

In order for the set-up to be physically meaningful, there must exist a hierarchy of scales.
In terms of the $r$ radial coordinate:\footnote{Being strictly precise, there
should be an extra scale $0 < r_q \ll r_h$ proportional to the quark masses 
if one wants to avoid the IR singularity of the zero temperature solution. 
The solutions presented in this section can be seen as the leading ones in 
$\frac{r_q}{r_h}
\sim \frac{m_q}{T}$. If one wants to use the results for phenomenological estimates of the QCD plasma, this should be a good approximation since $m_u, m_d \ll T$. Section \ref{S5massive} deals with the backreaction of massive quarks.
}
\be
r_h \ll r_* \ll r_a < r_{LP}\,\,.
\ee
The quantity $r_h$ sets the scale of the plasma temperature $\frac{r_h}{R^2}\sim \Lambda_{IR}\sim T$,
which is the scale at which we want to analyze the physics.
The point $r_{LP}$ is where the dilaton diverges, signaling a Landau pole in the dual theory.
However, as discussed in \cite{unquenchedmesons}, the string solution starts presenting
pathologies at a lower scale, which we denote by  $r_a$.
At this scale, which is fairly close to $r_{LP}$, the holographic $a$-function is singular and the utility of the solution for $r>r_a$ is doubtful.
Finally, $r_*$ sets an (arbitrary) UV cutoff scale $\frac{r_*}{R^2}\sim \Lambda_{UV}$.
The solution (\ref{finiteTsol}) will only be used for $r<r_*$.
In a Wilsonian sense of a renormalization group flow, the UV details
should not affect the IR physical predictions.
This feature is reflected in that physical quantities do not depend (up to suppressed contributions)
on $r_*$ or functions evaluated at that point, but only on IR parameters.
Even if the precise value of $r_*$ is arbitrary, we have to make sure that it is possible
to choose it such that it is well above the IR scale (so that the UV
completion 
only has negligible effects on the IR physics) and well below the pathological $r_a, r_{LP}$ scales
(so that the solution we use is meaningful and the expansions do not break down).
To this we turn now.

Let us start by  computing the hierarchy between $r_*$ and $r_{LP}$. Since at
$r_*$ we can approximate the solution by the supersymmetric one, we can read the
position of the Landau pole from (\ref{susysol}). If we insert 
the approximate relation between radial coordinates
$r \approx \sqrt{\alpha'} e^{\rho}$, we find:
\be
\frac{r_*}{r_{LP}}\approx e^{-\frac {1} {\epsilon_*}} \ll 1\,\,,
\ee
as long as $\epsilon_* \ll 1$.

Moreover, one has to make sure that the Taylor expansions (\ref{finiteTsol})
are valid in the region $r_h<r<r_*$. This of course requires $\epsilon_* \ll 1$,
but also that $\epsilon_* \left|\log\frac{r_h}{r_*}\right|\ll 1$ (notice that the absolute value of the
logarithm can be big because $r_h \ll r_*$). This means that
$\frac{r_h}{r_*} \gg e^{-\frac {1} {\epsilon_*}}$.
On the other hand, when we compute physical quantities in the upcoming sections, we always neglect
quantities suppressed as powers of $\frac{r_h}{r_*} \sim \frac{T}{\Lambda_{UV}}$.\footnote{For the thermodynamical quantities, we always neglect terms
suppressed, at least, as $\left(\frac{r_h}{r_*}\right)^4$, while for the jet quenching parameter
the neglected terms are of order $\frac{r_h}{r_*}$.
} This is the order of magnitude
of the corrections due to the eventual UV completion of the theory at $r_*$. One has to make sure that
the corrections in $\epsilon_*$ we are keeping are much larger than the neglected ones,
namely $\epsilon_* \gg \frac{r_h}{r_*}$. In summary, we have the 
following hierarchy of
parameters (in the following, in order to avoid overly messy expressions, we insert the value of $\epsilon_*$ for the
$X_5=S^5$ case, remembering that for a generic $X_5$, its value is given by
(\ref{epsstar})):
\be
e^{-\frac {1} {\epsilon_*}}\sim 
e^{-\frac {8\pi^2\,N_c} {\lambda_*\,N_f}}
\ll \frac{r_h}{r_*} \sim \frac{T}{\Lambda_{UV}}
\ll \epsilon_* \sim \frac{\lambda_*\,N_f}{8\pi^2\,N_c} \ll 1\,\,.
\ee
As long as $\epsilon_* \sim \frac{\lambda_*\,N_f}{8\pi^2\,N_c} \ll 1$, there always
exists a range of $r_*$ such that this inequality is satisfied. Since we will focus on the IR physics of the plasmas, at the scale set by their temperature, the actual physical constraint on the parameters will be
$
\frac{\lambda_h}{8\pi^2}\frac{N_f}{N_c} \ll 1\,\,,
$
which we have written in terms of  the coupling at the scale of the horizon,
$\lambda_h = \lambda_* (1 + O(\epsilon_*))$, see section \ref{thermosec} below.

On top of this, we have to make sure that the SUGRA+DBI+WZ action we are using is
valid. As usual, the suppression of
closed string loops requires
$N_c \gg  1$ whereas the suppression of $\alpha'$-corrections
is guaranteed by $\lambda_h \gg 1$. 
We have written the D7 worldvolume contribution to the action as a sum of $N_f$ single brane contributions.
This is justified if the typical energy of a string connecting two different branes is large
(in $\alpha'$ units). Since the branes are distributed on a space whose size is controlled
by $R \sim \lambda_h^\frac14 \sqrt{\alpha'}$, we again need $\lambda_h \gg 1$. 
The smearing approximation will be good if
 the distribution of D7-branes on the transverse space is dense, i.e. $N_f\gg 1$.
The discussion up to now is summarized in the following validity regime:
\be\label{validity}
N_c \gg  1\,\,,\qquad  \lambda_h \gg 1\,\,,\qquad N_f \gg 1 \,\,,\qquad 
\epsilon_h= \frac{\lambda_h}{8\pi^2}\frac{N_f}{N_c} \ll 1\,\, .
\ee
Finally, we want to find the regime of parameters in which the flavor corrections are not
only {\it valid} but are also the {\it leading} ones. With this aim,
we ought to demand that $\alpha'$-corrections to the supergravity action
(which typically scale as $\lambda_h^{-\frac32}$
due to first string corrections of the type $\alpha'^3 R^4$) are smaller than the flavor
ones, controlled by $\epsilon_h$, namely:\footnote{Notice that if one wants the second order flavor corrections that will be computed to also dominate over the curvature
ones, the more restrictive condition $\lambda_h^{-\frac32}\ll\epsilon_h^2$ is 
needed.}
\be
\lambda_h^{-\frac32} \ll \epsilon_h\,\,.
\label{leadingcond}
\ee
Demanding that corrections to the D7-branes contributions
(for instance curvature corrections to the worldvolume action itself or corrections produced by possible
modifications of the brane embeddings due to  curvature
corrections to the background metric)
are subleading does not impose any further restriction. The reason is that their contribution
is typically of order $\epsilon_h \lambda_h^{-c}$ for some $c>0$ which is always subleading
with respect to $\epsilon_h$ as long as (\ref{validity}) is satisfied.

Let us conclude this section with some comments on the stability of our perturbative non-extremal solutions.
A possible way to check for the latter is to consider worldvolume fluctuations of a D7-brane in the setup. If, as in our cases, the brane corresponds to massless flavors, the related quasi-normal modes on the unflavored background all have frequencies with a negative imaginary part of the order of the temperature, 
signaling stability \cite{mesonmelting}, \cite{myers}. 
This result cannot be changed in the flavored case when a perturbative 
expansion in $\epsilon_h$ is done. Thus, in our regime of approximations, stability with respect to those fluctuations is guaranteed.

\section{Thermodynamics of the solution}
\label{thermosec}
\setcounter{equation}{0}

In the previous section we have found a 
family of perturbative non extremal solutions which are regular at the horizon. 
They are dual to finite temperature flavored deformations of conformal theories, including ${\cal N}=4$ SYM.
The solution (\ref{finiteTsol}) is written in terms of the parameters 
$\epsilon_*,r_*$, which are defined at the UV scale. As already stressed, the physical quantities
must be expressed in terms of IR parameters. We thus define:
\be
\epsilon_h = \frac{\lambda_h\,Vol(X_3)}{16\pi\,Vol(X_5)}\frac{N_f}{N_c}\,\,,
\ee
where the subscript $h$ means that the quantities are evaluated at the horizon.
Thus, $\lambda_h$ is
naturally identified with the coupling at the scale of the plasma temperature.
We therefore have:
\be
\epsilon_h = \epsilon_* \frac{e^{\Phi_h}}{e^{\Phi_*}} = 
\epsilon_* + \epsilon_*^2 \log\frac{r_h}{r_*} + O(\epsilon_*^3)\,\,.
\label{epsirelation}
\ee
It is important to notice that, if we tune the temperature while keeping the UV parameters
fixed (namely, we change the temperature without changing the theory), then $\epsilon_h$
depends on the temperature. Since, as we will see below, $r_h$ is proportional to the temperature
(at leading order), we have:
\be
\frac{d\epsilon_h}{dT}=\frac{\epsilon_h^2}{T}+ O(\epsilon_h^3)\,\,,
\label{epsilonrun}
\ee
and $T(d\lambda_h/dT)=\epsilon_h\lambda_h$ at leading order. These relations reflect the running of the gauge coupling induced by the dynamical flavors. 

We now ask how the BH temperature and entropy density are related to the parameters of the solution. 
Let us start by computing the temperature requiring regularity of the euclideanized metric, by
identifying the temperature with the
inverse of the period of the euclideanized time.
A simple computation yields:
\be
T=\frac{2r_h}{2\pi R^2 \tilde S_{h}^4 \tilde F_{h}}=\frac{r_h}{\pi R^2}\left[1-\frac18 \epsilon_h
-\frac{13}{384}\epsilon_h^2 + O(\epsilon_h^3) \right]\,\,,
\label{temperature}
\ee
where we have inserted the values of $\tilde F, \tilde S$ at the horizon, which can
be read from (\ref{finiteTsol}) 
(neglecting terms suppressed
as powers of $\frac{r_h^4}{r_*^4}$):
\be
\tilde F_h= 1 - \frac{\epsilon_h}{24} + \frac{17}{1152} \epsilon_h^2 + O(\epsilon_h^3)\,\,,\qquad
\tilde S_h= 1 + \frac{\epsilon_h}{24} + \frac{1}{128} \epsilon_h^2+ O(\epsilon_h^3)\,\,.
\label{FSphi}
\ee
The entropy density $s$ is proportional to $A_8$, the volume at the horizon of the eight dimensional part of the space orthogonal to the $\hat t,r$ 
plane (where $\hat t$ is the Euclidean time), divided by the infinite constant volume of the 3d space directions
$V_3$. From the general form of the metric (\ref{deformedads5bh}) we get that:
\be
s=\frac{2\pi\,\,A_8}{\kappa_{(10)}^2\,V_3} =
\frac{r_h^3 R^2 \tilde S_h^4 \tilde F_h Vol(X_5)}{2^5 \pi^6 g_s^2 \alpha'^4}
=\frac{\pi^5}{2Vol(X_5)} N_c^2 \frac{r_h^3}{\pi^3 R^6}\left[1+\frac18 \epsilon_h
+\frac{19}{384}\epsilon_h^2 + O(\epsilon_h^3) \right]\,\,,
\ee
which in terms of the temperature reads:
\be
s=
\frac{\pi^5}{2Vol(X_5)} N_c^2 T^3 \left[1+\frac12 \epsilon_h
+\frac{7}{24}\epsilon_h^2 + O(\epsilon_h^3) \right]\,\,.
\label{entropy}
\ee
As for the other thermodynamic quantities which will follow, the leading term of this formula is the well-known unflavored result.
The $O(\epsilon_h)$ term was already calculated in \cite{myers} with the probe brane technique,
in the $X_5=S^5$ case. We re-obtain the result in a quite standard way, 
by computing the increase of the horizon area produced by the flavor branes. This can be considered
as a crosscheck of the validity of the whole construction.
Finally, the order $\epsilon_h^2$ term is a previously unknown contribution. 

The expression we have found in (\ref{entropy}),
shows that the relation $s= 2\pi^2 a_{FT}(T) T^3$, where $a_{FT}$ is the holographic $a$-charge of the $T=0$ theory, holds also in the flavored case, although only up to first order in $\epsilon_h$. Let us show this explicitly. The general formula for the holographic $a$-function of a flavored $T=0$ SQCD-like theory dual to the D3-D7 background  is given by $a_{FT}= 27(8\pi^5g_s^2\alpha'^4)^{-1} \beta^\frac32 H^\frac72 (\d_r H)^{-3}$ \cite{holaf}
where we have defined $H=R^4 r^6 Vol(X_5)^2 \tilde S_0^8 \tilde F_0^2$ and
$\beta = \frac{R^4}{r^4} \tilde S_0^8 \tilde F_0^2$ \cite{kkm}.
Inserting  (\ref{susymasslesssol}),
we get:
\be
a_{FT}=\frac{\pi^3 N_c^2}{4Vol(X_5)}\left[1 + \frac{\epsilon_*}{2} + \epsilon_*^2 \left( \frac16 + \frac12 \log\frac{r}{r_*}
+ \frac{5}{972}  \frac{r^8}{r_*^8} 
\right)+O(\epsilon_*^3) \right]\,\,.
\label{aFT}
\ee
This formula shows that the holographic central charge is constant up to first order in $\epsilon_*$. However, it starts running at second order, as a signal of the broken conformal
invariance. Defining $a_{FT}(T)= a_{FT}(r_h)$, we see that the claimed
relation with the entropy holds only at first order.

The ADM energy of the solution
can be straightforwardly computed (see appendix \ref{free_energy}). 
It yields the energy density of the plasma and, thus, it
allows us  to 
study the full thermodynamics:
\be
\varepsilon= \frac{E_{ADM}}{V_3}=\frac38 \frac{\pi^5}{Vol(X_5)} N_c^2 T^4
\left[1+\frac12 \epsilon_h(T) + \frac13 \epsilon_h(T)^2 +O(\epsilon_h(T)^3)\right]\,\,.
\label{ADMresult}
\ee
Again, terms suppressed as powers of $\frac{r_h}{r_*}$ have been neglected. Moreover,
since in the following derivatives with respect to $T$ are going to be taken, we find
it convenient to make explicit that $\epsilon_h$ depends on $T$ (see (\ref{epsilonrun})).
From the relation above we get immediately the heat capacity (density):
\be
{c_V} = \partial_T \varepsilon = \frac32 \frac{\pi^5}{Vol(X_5)}
N_c^2 T^3 \left[1+\frac12 \epsilon_h(T) +\frac{11}{24} \epsilon_h(T)^2+O(\epsilon_h(T)^3)\right] ~.
\ee
The free energy density, and so (minus) the pressure, reads:
\be\label{freeen}
\frac{F}{V_3} =-p = \varepsilon - T s= -\frac18 \frac{\pi^5}{Vol(X_5)}
N_c^2 T^4 \left[1+\frac12 \epsilon_h(T) +\frac16 \epsilon_h(T)^2+O(\epsilon_h(T)^3)\right] ~.
\ee
Notice that, consistently, this satisfies the relation $s=\partial_T p$ (where it is crucial
to take (\ref{epsilonrun}) into account).
This result is confirmed by the direct computation of $F$, which is relegated to appendix \ref{free_energy}.
Comparing (\ref{freeen}) to (\ref{aFT}), we find that $p=\frac{\pi^2}{2}a_{FT}(T)T^4$ up to second order.

The speed of sound $v_s$ is: 
\be
v_s^2 = \frac{s}{c_V} = \frac13 \left[1-\frac{1}{6} \epsilon_h(T)^2+O(\epsilon_h(T)^3)\right]~.
\label{vs2}
\ee
Note that the correction to the speed of sound 
only appears at second order. 
Instead, when quarks are massive, they break conformal symmetry at tree level and the correction to
the speed of sound is of first order in $\epsilon_h$ \cite{myers}.
It is also interesting to notice the sign of the correction, which is consistent with
the upper bound $v_s^2 \leq \frac13$ conjectured in \cite{Cherman:2009tw}
for a certain class of theories (see also \cite{Hohler:2009tv}). 
In \cite{Cherman:2009tw}, a heuristic motivation of the bound was given on the basis of asymptotic freedom, 
which it is not the case for the present theory.
As far as we know, (\ref{vs2}) is the first explicit check of this bound in the gravity dual
of a theory with positive $\beta$-function.

Analogous to the running of $a_{FT}$ (eq (\ref{aFT})), the
deviation from conformality in (\ref{vs2}) is a second order effect.
In fact, the solution provides a measure of the breaking of conformality at second order from the interaction measure:
\be
\frac{\varepsilon-3p}{T^4}=\frac{\pi^5 N_c^2}{16Vol(X_5)}\epsilon_h(T)^2\,\,.
\ee
The determination of this as well as other thermodynamic quantities on the lattice for QCD can be found in \cite{lattice}.
Of course, the running of observables in QCD and in our D3-D7 models are opposite.\footnote{In some sense, our solutions could represent a realization of the ``magnetic component'' of the QGP as described in \cite{Liao:2006ry}.}

Since we have not introduced into the action terms with higher derivatives of the metric (such as
curvature squared terms), the usual theorems apply and the fundamental matter does not affect the
result $\frac{\eta}{s}=\frac{1}{4\pi}$  \cite{Kovtun:2004de}. We can thus read the value of the shear viscosity trivially by dividing (\ref{entropy}) by $4\pi$.
Again, while the first order value was already calculated in \cite{Mateos:2006yd}, the second order result is new.

Concerning the bulk viscosity, if our thermal gauge theories saturate order by order the bound proposed in \cite{buchel}:
\be
\frac{\zeta}{\eta}\geq 2\left(\frac13 - v_s^2\right)\, ,
\label{bbound}
\ee
then $\zeta$ would still be zero up to first order. On the other hand, at second order we would obtain a non-trivial result:\footnote{The fact that the bulk viscosity should be seen as a second order effect was already observed in \cite{Mateos:2006yd}.}
\be
\zeta=\frac{\pi^4}{72 Vol(X_5)}N_c^2T^3 \left[\epsilon_h(T)^2+O(\epsilon_h(T)^3)\right]\, .
\ee
Adapting the reasoning in \cite{gubserbulk} to our case, one can find indications that the bound is indeed saturated, provided some (reasonable) assumptions are fulfilled. It would be very interesting to perform the precise holographic calculation of $\zeta/\eta$ to check whether this is actually the case.

\section{Energy loss of  partons in D3-D7 plasmas}
\setcounter{equation}{0}
\label{jetsection}
Let us now consider the energy loss of partons in the quark-gluon plasmas dual to the non-extremal D3-D7 solutions we have found in section \ref{sec:finiteT}.

The holographic study of energy loss has turned out to be quite relevant from a phenomenological point of view. This is because the real-world QCD quark-gluon plasma, whose properties are being studied at RHIC and will be investigated at the LHC, appears to be strongly coupled \cite{shur}. Moreover, the phenomenon of jet quenching observed at RHIC demands for a very efficient mechanism of energy loss. There are two main ways to account for this phenomenon in the stringy picture.\footnote{See also \cite{Gubser:2008as}.}

On one hand,  it is natural to try to model jet quenching as a result of the bremsstrahlung that occurs through the strong interactions of the parton probe with the quarks and gluons in the plasma. In perturbation theory, this mechanism is effectively captured by a transport coefficient termed $\hat q$, the jet-quenching parameter \cite{Baier}.
At very high energy, and using  the eikonal approximation, the authors of \cite{liu} found a non-perturbative prescription for calculating $\hat q$ 
as the coefficient of $L^2$ in an almost  light-like Wilson loop with dimensions $L^{-}\gg L$. Being non-perturbative it can be implemented in a string theoretic framework and this was done for the first time in \cite{liu}. 
Following this prescription,\footnote{In \cite{Marquet:2009eq} a different approach to the same problem has been recently proposed.} we will compute $\hat q$ for our backreacted background in section \ref{sec: jetqu}.

On the other hand, at strong coupling  the energy loss of a probe passing through a medium can be modeled entirely within a string theoretic framework  \cite{herzog1,gubserdrag}.
A parton of velocity $v$
is described by a macroscopic string attached to a probe flavor brane.
The string is dragged by a constant force $f$ which keeps the velocity fixed.
The drag force transfers energy and momentum to the parton, which are lost in the plasma at a constant rate.
The drag coefficient $\mu$, which measures the energy loss, is then calculated from the equation $f=\mu p$, where $p$ is the parton momentum.
In section \ref{sec: drag}, we will deal with this drag
force in the flavored background.

It is worth noticing that the computations we will make with the Nambu-Goto action receive 
stringy corrections
of order $\lambda^{-\frac12}$  \cite{Forste:1999qn}. If we want them to be subleading with respect to the
flavor corrections, we need to require $\lambda^{-\frac12} \ll \epsilon_h$,
which is more restrictive than (\ref{leadingcond}).

\subsection{The jet quenching parameter}
\label{sec: jetqu}

Let us compute the jet quenching parameter $\hat q$ following the prescription of
\cite{liu}. Taking the generic formula in \cite{aredmas}\footnote{We took into account a different factor of $\sqrt2$ between the definition of \cite{liu} and of \cite{aredmas}.} (and cutting the integral at
$r_*$), 
we can write:
\be
\hat q^{-1}= \pi\, \alpha' \int_{r_h}^{r_*} e^{-\frac{\Phi}{2}}
\frac{\sqrt{g_{rr}}}{g_{xx}\sqrt{g_{xx}+g_{tt}}}dr
= \frac{R^4\pi\, \alpha'}{r_h^2} e^{-\frac{\Phi_h}{2}}\int_{r_h}^{r_*} e^{-\frac{(\Phi-\Phi_h)}{2}}
\frac{\tilde S^4 \tilde F}{\sqrt{r^4-r_h^4}}dr
\,\,.
\ee
The dilaton enters the formula because we are considering the Einstein frame metric.
We have extracted $e^{-\frac{\Phi_h}{2}}$ because we want to factor out the physical IR  parameter
$\sqrt{\lambda_h} = e^{\frac{\Phi_h}{2}} \sqrt{4\pi \,g_s \,N_c}=e^{\frac{\Phi_h}{2}}\frac{R^2}{\alpha'}
\frac{\sqrt{Vol(X_5)}}{\pi^\frac32}$.
We can now insert (\ref{finiteTsol}) and
write everything in terms of $\epsilon_h$ rather than $\epsilon_*$. Finally, performing a change of
variable $\varrho=\frac{r}{r_h}$, we can write the inverse of the jet quenching parameter as:
\be
\hat q^{-1}= \frac{\pi\,R^6}{r_h^3 \sqrt{\lambda_h}}\frac{\sqrt{Vol(X_5)}}{\pi^\frac32}
\left[  I_0 + I_1 \epsilon_h + I_2 \epsilon_h^2 + O(\epsilon_h^3) \right]\,\,.
\ee
The integrals that appear at each order are:
\bear
I_0 &=& \int_1^\infty \frac{1}{\sqrt{\varrho^4 -1}}d\varrho= \frac{\sqrt\pi\, \Gamma(\frac54)}{\Gamma(\frac34)}\,\,,\rc
I_1 &=& \int_1^\infty \frac{1-4\log\varrho}{8\sqrt{\varrho^4 -1}}d\varrho= \frac{\sqrt\pi\, \Gamma(\frac54)}{\Gamma(\frac34)}\,
\frac{1-\pi}{8}\,\,,\rc
I_2 &=& \int_1^\infty \frac{19-8\log\varrho-48\log^2\varrho - 12\, Li_2(1-\varrho^{-4})
}{384\sqrt{\varrho^4 -1}}d\varrho=\rc
&=& \frac{\sqrt\pi\, \Gamma(\frac54)}{\Gamma(\frac34)}\,
\frac{1}{384}\left[19-48 {\cal C}-\pi(2+3\pi)-8\ {}_4 F_3\left(
1,1,1,\frac32;\frac74,2,2;1\right)
\right]\,\,,
\label{3integrals}
\eear
where ${\cal C}\sim 0.91597$ is the Catalan constant. The upper limit of the integral has been taken 
to infinity instead of $\varrho_*=\frac{r_*}{r_h}$ since 
all the integrands are of order $\varrho^{-2}$ at large $\varrho$ and, therefore,
$\int^\infty - \int^{\varrho_*} = O(\varrho_*^{-1}) = O(\frac{r_h}{r_*})$,
which, as usual, we disregard.
The jet quenching parameter in terms of 
gauge theory quantities reads:
\be
\hat q=\frac{\pi^3\sqrt{\lambda_h}\Gamma(\frac34)}{\sqrt{Vol(X_5)}\,\Gamma(\frac54)}T^3
\left[1 + \frac18(2+\pi) \epsilon_h + \gamma\, \epsilon_h^2+ O(\epsilon_h^3)
\right]\,\,,
\label{jetq}
\ee
where we have introduced a constant $\gamma$:
\be
\gamma= \frac{11}{96}+\frac{\pi}{48} + \frac{3\pi^2}{128} + \frac18 {\cal C}+
\frac{1}{48}\ {}_4 F_3\left(
1,1,1,\frac32;\frac74,2,2;1\right)\approx 0.5565\,\,.
\ee
\subsubsection{Possible implications for phenomenology}
\label{sectioncompare}

We can now discuss some physics coming from  (\ref{jetq}). In \cite{lrw2}, it
was shown that for a class of
theories with only adjoint (and bifundamental) fields with gravity duals, one can write:
\be
\hat q = c \,\sqrt{\lambda_h} \sqrt{\frac{s}{N_c^2}}T^\frac32\,\,,\qquad\qquad
c=\sqrt{2\pi}\frac{\Gamma(\frac34)}{\Gamma(\frac54)}\,\,.
\label{jq1}
\ee
It was argued that deviations from this formula could come from having fundamental fields or from
non-conformality of the theory. In our case we actually deal with both of these effects.
As far as we know, this is the first set-up in which one can directly test the effect of fundamental
fields on the jet quenching parameter in a framework completely under control.
The expression (\ref{jq1}) is modified to:
\be
\hat q = c \,\sqrt{\lambda_h} \sqrt{\frac{s}{N_c^2}}T^\frac32
\left[1+\frac{\pi}{8}\epsilon_h+(\gamma-\frac{11}{96}-\frac{\pi}{32})\epsilon_h^2+
O(\epsilon_h^3) \right]\,\,.
\ee
Thus, from the point of view of this formula, $\hat q$ increases if $N_f>0$.
At first sight, one could think that this enhancement of the jet quenching parameter is
a trivial effect due to the increase of degrees of freedom of the plasma because of the
addition of new fields. 
This is a  naive  conclusion since, as remarked in \cite{liu}, the expression (\ref{jq1}) 
shows that, contrary to previous expectation, $\hat q$ is not linked to the number of degrees
of freedom. However, it is instructive to make the following gedanken experiment:\footnote{We
thank David Mateos for suggesting this comparison.} imagine we have a flavor-\emph{less} plasma at
temperature $T$, entropy density $s$ and with $N_{c,1}$ colors. One can think of replacing some
of the adjoints by fundamentals, while keeping the total number of degrees of freedom
(namely, the entropy density $s$) fixed. We will also maintain fixed the temperature
$T$ and the coupling constant 
$\alpha_s = g_{YM}^2/(4\pi)$ at the scale set by $T$.\footnote{Fixing $\lambda_h$ instead of $\alpha_s$ produces the same result. Fixing $N_c$ and $\epsilon_h$ gives instead the opposite behavior, but it seems a pathological choice, since it does not include the unflavored limit $\epsilon_h\rightarrow 0$.} We thus compare the theory with $N_{c,1}$ colors and no flavors to a theory with 
$N_{c,2}\ (<N_{c,1})$ colors and $N_f$ flavors. Concretely, keeping $T$, $s$ fixed requires
$N_{c,2}=N_{c,1}(1-\frac14 \epsilon_h-\frac{5}{96}\epsilon_h^2)$, where we have used (\ref{entropy}). 
For the flavored theory, but written in terms of $N_{c,1}$ (the corresponding number of
colors of the unflavored theory), we would have:
\be\label{qcompare}
\hat q = c \,\sqrt{N_{c,1} \,\,g_{YM}^2} \sqrt{\frac{s}{N_{c,1}^2}}T^\frac32
\left[1+\frac{1+\pi}{8}\epsilon_h+(\gamma-\frac{25}{384}-\frac{\pi}{64})\epsilon_h^2 +
O(\epsilon_h^3) \right]\,\,.
\ee
The positive sign of the correction terms shows that the flavors enhance the jet quenching.
This conclusion is confirmed in the alternative comparison scheme proposed in \cite{Gubser:2006qh}, in which the energy density and the force between external quarks are kept fixed, while $T$ and $\lambda_h$ are varied.\footnote{To be more specific and working at first order in $\epsilon_h$: $N_c$ is kept fixed, so from the matching of $\varepsilon$ one gets $T_2=T_1(1-\epsilon_h/8)$, so $\hat q_2/\hat q_1 =\sqrt{\lambda_{h,2}/\lambda_{h,1}}[1+(\pi-1)\epsilon_h/8]$. The coupling $\lambda_h$ is adjusted in such a way that the force between two external quarks at the screening length of the first plasma, $\alpha_{qq}=3 L_{c,1}^2 V'(L_{c,1})/4$, is kept fixed \cite{Gubser:2006qh}. By numerically computing the potential $V(L)$ from the Wilson loop in the two plasmas and equating the two forces $\alpha_{qq}$, one straightforwardly finds that $\lambda_{h,2}>\lambda_{h,1}$ and so $\hat q_2>\hat q_1$.}
The fact that the jet quenching is enhanced in all of  these comparison schemes (we found no reasonable counter-examples) seems to indicate that the result is quite robust.

Interestingly,  the enhancement of the jet quenching by flavors  was already observed in \cite{noncrit} in the context of a non-critical five-dimensional string dual of a flavored plasma. Of course, in that model the string corrections are not under control, while in the present case the result is completely trustworthy. 

Let us close this section by making a numerical estimate of the jet quenching parameter, by
inserting quantities relevant at RHIC, even if it is not clear which is the best way
to extrapolate the results from the theories here discussed when giving estimates for QCD.
As a first step, in order to compare with the unflavored ${\cal N}=4$ SYM result in \cite{liu}, let us pick  $Vol(X_5)=Vol(S^5)=\pi^3$ and extrapolate our result to the realistic regime (which is not included in the regime of validity of any holographic model) where $\alpha_s\sim 1/2$ and  $N_c=3$, i.e. $\lambda_h\sim 6\pi$.
Then we would have $\epsilon_h\sim\frac{1}{4\pi}N_f\sim 0.24$ for $N_f=3$.
From (\ref{jetq}) we read that the correction with respect to the ${\cal N}=4$ SYM result
corresponds to an increase of about $20\%$ for $N_f=3$. As an example, at $T=300$ MeV we would get $\hat q \sim 5.3$ (Gev)$^2$/fm, to be compared with the value $\hat q \sim 4.5$ (Gev)$^2$/fm of the unflavored plasma \cite{liu}. The flavored result would be right in the ballpark of RHIC values, i.e. $\hat q \sim 5-15$ (Gev)$^2$/fm. 

This comparison with QCD involves two theories with a different number of degrees of freedom.
Should we compare the two theories at fixed temperature, entropy density and $\alpha_s$, as discussed above, we would get a smaller $\hat q$. Let us take
$s_{QCD}/T^3\sim 17.5$ at $T=300$ MeV from \cite{lattice} (figure 10). Then, using (\ref{entropy}),
and inserting
$N_f=3$ and $\epsilon_h=\frac{3}{4\pi}$, we get 
a smaller coupling $\lambda_h\sim 11.1$. This would imply, from (\ref{jetq}), that $\hat q\sim 4.1$ (Gev)$^2$/fm. 
In the same scheme, in the flavorless ${\cal N}=4$ SYM limit we would get $\hat q\sim 3.5$ (Gev)$^2$/fm.

Let us note that the interaction measure in the latter scheme is rather small, $(\varepsilon-3p)/T^4=N_c^2\pi^2 \epsilon_h^{2}/16 \sim 0.11$. 
In lattice QCD it is also known to be small at temperatures above $2T_c\sim $ 400 MeV, where the theory is nearly conformal, but not that much.
For example, from figure 4 in \cite{lattice}, it is around 1 for temperatures around 500 MeV.
Of course, in QCD $(\varepsilon-3p)/T^4$ has a large contribution also at the flavorless level, being proportional to the trace anomaly.

\subsection{The drag force}
\label{sec: drag}

Let us now consider a heavy quark
moving through our D3-D7 plasmas and compute the drag force it experiences, following the general procedure described in \cite{herzog1,gubserdrag,herzog2}. We consider
a simple string configuration representing a test quark moving in a
given spatial direction $x$: $t = \hat\tau,\ r = \hat\sigma,\ x = x(\hat\sigma,\hat\tau)$.
In particular, we will just discuss the stationary string configuration (an open string with an
extremum attached to a probe D7-brane at $r=r^*$) corresponding to a quark
which moves at constant velocity, such that the energy loss due to friction with the
medium is compensated by an external force. 
This is achieved by setting
$x(\hat\sigma,\hat\tau) = r(\hat\sigma) +v\,\hat\tau$. 
We have to analyze the 
Nambu--Goto action in
the background metric $ds_{str}^2 = e^{\Phi/2} ds^2$ with $ds^2$ given in (\ref{deformedads5bh}).
We can use general results from \cite{herzog2} and define $C$ as the constant 
determined from the equation $g_{xx}(r_c)g_{tt}(r_c)+C^2=0$ with the point $r_c$ given by
$g_{tt}(r_c)+g_{xx}(r_c)v^2=0$, namely $r_c=r_h(1-v^2)^{-\frac14}$.
Then, the rate of momentum transferred  to the medium is given by \cite{herzog2}:
\be
\frac{dp}{dt} = -\frac{1}{2\pi\alpha'}C = -
\frac{r_h^2}{2\pi\alpha'\,R^2}e^\frac{\Phi(r_c)}{2}\frac{v}{\sqrt{1-v^2}}=-
\mu\,M_{kin}\,\frac{v}{\sqrt{1-v^2}}\,\,,
\label{drag1}
\ee
where we have introduced notation from \cite{herzog1}:
a friction parameter $\mu$ such that $\frac{dp}{dt}=-\mu\,p$ and a
kinematical mass $M_{kin}$ such that $p=M_{kin}\frac{v}{\sqrt{1-v^2}}$.
From (\ref{drag1}), using (\ref{simpleh}), (\ref{finiteTsol}), (\ref{epsirelation}), (\ref{temperature}), 
we find:
\bear\label{muemme}
\mu\,M_{kin} &=& \frac{\pi^{5/2}}{2} \frac{\sqrt{\lambda_h}}{\sqrt{Vol(X_5)}}\, T^2 \left[
1+\frac18 (2-\log(1-v^2)) \epsilon_h +
\right. \\
&& \left.+
\frac{1}{384}\left[44-20 \log(1-v^2)
+9 \log ^2(1-v^2)+12 Li_2(v^2) \right] \epsilon_h^2 + O(\epsilon_h^3)\right]\,\,.\nonumber
\eear
As in section \ref{sec: jetqu}, the energy loss (at fixed $v$) is enhanced by the presence of fundamental matter, also in the different comparison schemes described in section \ref{sectioncompare}. The quantity $\mu\,M_{kin}$ grows when increasing the velocity. From (\ref{muemme}), formally,
it would diverge as $v\to 1$. However, 
(\ref{muemme}) is not applicable in that limit since 
we have to require $\epsilon_h \log(1-v^2) \ll 1$ for the expansions to be valid.

\section{Deforming $AdS_5 \times S^5$ with massive flavors}
\label{S5massive}
\setcounter{equation}{0}

In this section, we will write down the generalization of equations (\ref{bhSeqsmassless}), (\ref{constraintmassless}) for the
case in which the backreacting flavors are massive (all with the same modulus of the mass).
The search for solutions and the study of the physics is left for future work.
Some technical details associated to this section are relegated to appendix \ref{unquenchedkk}.

The extra complication we face when the quarks are massive is that the D7-brane embeddings are in this case non-trivial and, in fact, have to be studied
numerically (when $T>0$, even if backreaction is not taken into account). 
Moreover, this case cannot be studied with the same generality as the massless one, 
since the mentioned embedding equations do depend on the metric of the K\"ahler-Einstein space.
We will focus in the case where the set-up is a flavor deformation of the $AdS_5 \times S^5$ black hole.
Thus, the K\"ahler-Einstein space is $CP^2$ and the ten dimensional metric we consider is 
(\ref{10dmetric}) with:
\bear
ds_{CP^2}^2&=&\frac14 d\chi^2+
\frac14 \cos^2 \frac{\chi}{2} (d\theta^2 +
\sin^2 \theta d\varphi^2) 
+ \frac{1}{4} \cos^2 \frac{\chi}{2} \sin^2 \frac{\chi}{2}(d\psi + \cos \theta d\varphi)^2\,\,,\rc
A_{CP^2}&=& \frac12\cos^2 \frac{\chi}{2}(d\psi + \cos \theta d\varphi)\,\,.
\label{cp2metric}
\eear
The range of the angles is $0\leq \chi, \theta \leq \pi$, 
$0\leq \varphi, \tau < 2\pi$, $0\leq \psi< 4 \pi$.
The $F_{(5)}$ RR field strength takes the same form as in the massless case, but $F_{(1)}$ picks up
a dependence on the radial coordinate:
\bear
F_{(5)} = Q_c\,(1\,+\,*)\varepsilon(S^5)\,\,,\qquad
F_{(1)} = Q_f\, p(\sigma)\,(d\tau + A_{CP^2})\,\,,
\label{massiveforms}
\eear
where $p(\sigma)$ is a function which depends on the brane embeddings. It has to vanish at scales smaller
than the quark masses and to asymptote to 1 at energy scales much larger than the quark masses.

First of all, we have to write down the equation that determines the embedding of a D7-brane in this background.
Let us take as  a ``fiducial brane'' one wrapping $\theta,\varphi,\psi$, with $\chi$ a function of $\sigma$ and
situated at fixed $\tau$. The rest of the family of embeddings needed for the smearing are obtained from
the fiducial one by acting with the symmetries of the internal space, see appendix \ref{unquenchedkk} for a
discussion. 
The DBI action for this fiducial brane reads:
\be
S_{DBI} = -\frac{T_7}{8} \int d^8x\, e^{\Phi}b \sin\theta S^6F^2\cos^3 
\left(\frac{\chi_{wv}}{2}\right)
\sqrt{\cos^2 \left(\frac{\chi_{wv}}{2}\right)+\frac{S^2}{F^2}\sin^2 \left(\frac{\chi_{wv}}{2}\right)}
\sqrt{1+\frac{(\d_\sigma\chi_{wv})^2}{4b \,S^6\,F^2 }}\,\,,
\label{DBIgeneric}
\ee
where $\chi_{wv}$ is the function of $\sigma$ that determines the brane embedding. The 8d 
integral is taken along the Minkowski directions, $\theta,\varphi,\psi$ and $\sigma$.
There is also a WZ term due to the coupling to the background $C_{(8)}$. The $C_{(8)}$ potential can
be computed from $dC_{(8)} = e^{2\Phi} ({}^*F_{(1)})$.
The WZ piece of the action is:
\be
S_{WZ} = - \frac{T_7}{32}Q_f \int d^8x \sin\theta p(\sigma) b \,e^{2\Phi} S^8 \cos^4
\left(\frac{\chi_{wv}}{2}\right)\,\,.
\label{WZ3}
\ee
The equation of motion for $\chi_{wv}$ follows from the action $S_{wv}=S_{DBI}+S_{WZ}$:
\bear\label{chiwveq}
0&=&\frac12 \d_\sigma\left(e^\Phi \cos^3 \frac{\chi_{wv}}{2}\,\frac{\Xi_1}{\Xi_2}(\d_\sigma \chi_{wv})
\right)+\\
&+&e^\Phi b \, S^6 F^2 \cos^2 \frac{\chi_{wv}}{2}\sin \frac{\chi_{wv}}{2}
\left(3\Xi_1 \Xi_2 + \cos^2 \frac{\chi_{wv}}{2} (1-\frac{S^2}{F^2}) \frac{\Xi_2}{\Xi_1}
+Q_f e^\Phi \frac{S^2}{F^2}\cos \frac{\chi_{wv}}{2} p(\sigma)
\right)\,\,.\nonumber
\eear
In order to abbreviate the notation, we have introduced the quantities:
\be
\Xi_1 \equiv \sqrt{\cos^2 \left(\frac{\chi_{wv}}{2}\right)+\frac{S^2}{F^2}\sin^2 \left(\frac{\chi_{wv}}{2}\right)}\,\,,
\qquad\quad
\Xi_2\equiv \sqrt{1+\frac{(\d_\sigma\chi_{wv})^2}{4b \,S^6\,F^2 }}\,\,.
\ee
The next step is to write down the one-dimensional effective action for the closed string fields,
namely the dilaton and the functions that enter the ansatz for the metric. This is similar to
(\ref{effeclagr}) but one has to take into account the $\sigma$-dependence of $F_{(1)}$ and
that the DBI term is $N_f$ times 
(\ref{DBIgeneric}). As before, the WZ term does not contribute because it does
not depend on the metric and the dilaton. We get:
\bear
S_{eff}&=&\frac{\pi^3 V_{1,3}}{2\kappa_{10}^2}\int d\sigma \left(
-\frac12\frac{(\d_\sigma h)^2}{h^2} +12\frac{(\d_\sigma S)^2}{S^2} + 8 \frac{(\d_\sigma F)(\d_\sigma S)}{F\,S}-
\frac12 (\d_\sigma \Phi)^2+\right.\rc
&+&\left.
\frac{(\d_\sigma b)}{2b}\left( \frac{(\d_\sigma h)}{h}+ 8 \frac{(\d_\sigma S)}{S}+ 2 \frac{(\d_\sigma F)}{F} 
\right)+
24 b\,F^2\,S^6 - 4 b\,F^4\,S^4 +
\right. 
\rc
&-&\left.
\frac12 Q_c^2 \frac{b}{h^2}  -\frac12 Q_f^2 p(\sigma)^2 e^{2\Phi}
b\,S^8\, 
-
4Q_f \cos^3 \left(\frac{\chi_{wv}}{2}\right) \Xi_1 \Xi_2
e^{\Phi} \,b\, S^6\,F^2 \right)\,\,.
\label{effeclagrmassive}
\eear
One can readily compute the equations of motion from this Lagrangian.
An important point to note is that $\chi_{wv}$ is not a field in this lagrangian. Its equation
of motion (\ref{chiwveq}) was derived from the worldvolume action.
We find the following set of Euler-Lagrange equations:
\bear
\partial_\sigma^2(\log b)&=&0\,\,,\rc
\partial_\sigma^2(\log h)&=&-Q_c^2 \frac{b}{h^2}\,\,,\rc
\partial_\sigma^2(\log S)&=& -2 b F^4 S^4 + 6 b F^2 S^6 -\frac12 Q_f \,e^\Phi b\,F^2\,S^6\,
\cos^3 \frac{\chi_{wv}}{2}\left(\frac{\Xi_1}{\Xi_2} + \cos^2 \frac{\chi_{wv}}{2}
\frac{\Xi_2}{\Xi_1}\right)\,\,,
\rc
\partial_\sigma^2(\log F)&=& 4b\,F^4 S^4 - \frac{Q_f^2}{2}e^{2\Phi}b\,S^8p(\sigma)^2
-2\frac{\Xi_2}{\Xi_1} e^\Phi Q_f\,b\,S^8 \cos^3 \frac{\chi_{wv}}{2} \sin^2 \frac{\chi_{wv}}{2}\,\,,
\rc
\partial_\sigma^2\Phi&=& 
Q_f^2\,e^{2\Phi}\,b\,S^8 p(\sigma)^2 + 4 Q_f e^\Phi\,b\,S^6 F^2\,\cos^3 \frac{\chi_{wv}}{2}\,\Xi_1\Xi_2\,\,.
\label{bhSeqsmassive}
\eear
The first order constraint, generalizing (\ref{constraintmassless}), reads:
\bear
0&=&-\frac12\frac{(\d_\sigma h)^2}{h^2} +12\frac{(\d_\sigma S)^2}{S^2} + 8 \frac{(\d_\sigma F)(\d_\sigma S)}{F\,S}-
\frac12 (\d_\sigma \Phi)^2+\rc
&+&
\frac{(\d_\sigma b)}{2b}\left( \frac{(\d_\sigma h)}{h}+ 8 \frac{(\d_\sigma S)}{S}+ 2 \frac{(\d_\sigma F)}{F} 
\right)-24 b\,F^2\,S^6 + 4 b\,F^4\,S^4+
\rc
&+&
\frac12 Q_c^2 \frac{b}{h^2}  +\frac12 Q_f^2 e^{2\Phi}
b\,S^8\, p(\sigma)^2
+
4Q_f
e^{\Phi} \,b\, S^6\,F^2 \cos^3 \frac{\chi_{wv}}{2}\,\frac{\Xi_1}{\Xi_2}\,\,.
\label{constraintmassive}
\eear
The only ingredient left to know is the precise expression for $p(\sigma)$.
The answer is given in appendix \ref{unquenchedkk}:
\be
p(\sigma)=\cos^4\frac{\chi_{wv}}{2}\,\,.
\label{pofsigma}
\ee
The system of equations (\ref{chiwveq}), (\ref{bhSeqsmassive}), (\ref{constraintmassive}),
together with (\ref{pofsigma}) defines the solution for a finite temperature 
D3-D7 background where the quarks are massive (all of them with the same modulus of the mass).
If at some value $\sigma=\sigma_q$ above the horizon $\chi_{wv}$ reaches the value
$\chi_{wv}=\pi$, it means that the flavor branes only have support for
$\sigma\geq\sigma_q$. In that case, it has to be understood that equations
(\ref{chiwveq}), (\ref{bhSeqsmassive}), (\ref{constraintmassive}),
(\ref{pofsigma}) are valid for
$\sigma>\sigma_q$ whereas for $\sigma<\sigma_q$, one has the unflavored system of equations
(namely one should substitute $Q_f=0$ for $\sigma<\sigma_q$ in all the equations). 
At $\sigma=\sigma_q$, one has to impose appropriate
matching conditions. Detailed discussions of this feature in various supersymmetric cases
can be found in \cite{bcnp,fkw1,fkw2,Bigazzi:2008qq}.

By deriving the constraint (\ref{constraintmassive})
with respect to $\sigma$ and inserting (\ref{chiwveq}),
(\ref{bhSeqsmassive}), (\ref{pofsigma}), one obtains an identity, proving that the system of equations is self-consistent.
This can be considered a crosscheck for the whole procedure.
The massless limit is given by $\chi_{wv}=0$, $\Xi_1=\Xi_2=1$, and one recovers the
set-up of section \ref{sec: massless}.

\subsection{The supersymmetric case}

We now consider the zero temperature solution of the system with massive flavors.
One can check that the following first order 
equations solve
(\ref{chiwveq}), (\ref{bhSeqsmassive}), (\ref{constraintmassive}):
\bear
&&b=1\,\,,\qquad\qquad\
\d_\sigma h=-Q_c\,\,,\qquad
\qquad
\d_\sigma F= S^4 F \left( 3-2\frac{F^2}{S^2} - \frac{Q_f}{2} e^\Phi \cos^4\frac{\chi_{wv}}{2}\right)\,\,,\qquad\rc
&&\d_\sigma S= S^3 F^2\,\,,\qquad\qquad
\d_\sigma \chi_{wv} = -2S^4 \tan \frac{\chi_{wv}}{2}\,\,,\qquad\quad
\d_\sigma \Phi= Q_f\,S^4 e^\Phi \cos^4\frac{\chi_{wv}}{2}\,\,.\qquad
\label{massiveBPS}
\eear
These are BPS conditions, generalizations of (\ref{masslessBPS}), which determine the
supersymmetric solutions.
It is convenient to
define a $\rho$ coordinate as $d\rho=S^4d\sigma$.
In the $\rho$ coordinate, the embedding is just $\sin\frac{\chi_{wv}}{2} = e^{\rho_q-\rho}$, where $\rho_q$ is an integration constant related to the bare quark masses.
We can write explicit expressions for $S,F,\Phi$, for $\rho>\rho_q$:\footnote{In integrating $S,F, \Phi$, we have fixed integration constants as in the massless case. Moreover we have required $S|_{\rho=\rho_q} = F|_{\rho=\rho_q}$ to ensure regularity in the IR (at $\rho=-\infty$) as explained in \cite{fkw1}.}
\bear
S&=&\alpha'^{\frac12}\, e^\rho\,\left(1+\epsilon_* (\frac16 +\rho_*-\rho-\frac16 e^{6\rho_q-6\rho}
-\frac32 e^{2\rho_q-2\rho} + \frac34 e^{4\rho_q-4\rho}-\frac14 e^{4\rho_q-4\rho_*}+ e^{2\rho_q-2\rho_*}
)\right)^\frac16
\rc
F&=&\alpha'^{\frac12}\, e^\rho\,\frac{\left(1+\epsilon_* (\rho_*-\rho-e^{2\rho_q-2\rho}+
\frac14 e^{4\rho_q-4\rho}
+e^{2\rho_q-2\rho_*}-
\frac14 e^{4\rho_q-4\rho_*}
)\right)^\frac12}
{\left(1+\epsilon_* (\frac16 +\rho_*-\rho-\frac16 e^{6\rho_q-6\rho}
-\frac32 e^{2\rho_q-2\rho} + \frac34 e^{4\rho_q-4\rho}-\frac14 e^{4\rho_q-4\rho_*}+ e^{2\rho_q-2\rho_*}
)\right)^{\frac13}}
\rc
\Phi&=& \Phi_* -\log(1+\epsilon_*\, (\rho_*-\rho-e^{2\rho_q-2\rho}+
\frac14 e^{4\rho_q-4\rho}
+e^{2\rho_q-2\rho_*}-
\frac14 e^{4\rho_q-4\rho_*}))\,\,.
\label{susysolmassive}
\eear
Setting $\rho_q \to -\infty$ one recovers the massless
solution (\ref{susysol}). 

We still have to write the solution for $\rho<\rho_q$. The flavor branes do not reach this region and the
equations of motion for the background are (\ref{massiveBPS}) with $Q_f=0$.
Thus, the dilaton is constant and, by continuity, it has the value that can be read from 
(\ref{susysolmassive}) inserting $\rho=\rho_q$:
\be
\Phi_{IR}=\Phi_q = \Phi_* -\log(1+\epsilon_*\, (\rho_*-\rho_q-
\frac34 
+e^{2\rho_q-2\rho_*}-
\frac14 e^{4\rho_q-4\rho_*}))\,\,.
\ee
The functions $S$ and $F$ are equal:
\be
S=F=\alpha'^{\frac12}\, e^\rho e^{-\frac16(\Phi_{IR}-\Phi_*)}\,\,,\qquad\qquad
(\rho<\rho_q)\,\,.
\ee
We defer a more detailed analysis of these solutions and of their generalizations to finite
temperature to future work.

\section{Summary}
\setcounter{equation}{0}
\label{sec: discussion}

In this paper we have
presented a backreacted supergravity solution dual to a flavored version of the ${\cal N}=4$ SYM plasma, with massless flavors in the group $U(1)^{N_f}$.
Our construction is also valid for black holes on $AdS_5 \times X_5$ backgrounds, with $X_5$ a generic Sasaki-Einstein space. In all the cases the $T=0$ solutions preserve
${\cal N}=1$ supersymmetry in 4d.

The solutions we provide are perturbative, like the one presented in \cite{thermoks} for the unflavored thermal conifold theory.
The relevant expansion parameter here,  $\epsilon_h\sim \lambda_h N_f/N_c$, weighs 
the internal flavor loop contributions to the unflavored field theory background. In the range of physical parameters associated to
RHIC, we can reasonably expect values of $\epsilon_h\sim 0.24$. On the other hand, this makes it a nice expansion parameter, and
keeping corrections up to order $\epsilon_h^2$ yields a very good approximation to the full result. This is precisely the order at which 
we are able to provide a solution in the present paper.
Moreover, leading order $\epsilon_h$ corrections are already of about $20\%$ which are clearly dominant over curvature corrections. 
These come with a  $\lambda_h^{-3/2}$ for string corrections to the supergravity lagrangian
and with $\epsilon_h \lambda_h^{-1}$ for corrections coming from the DBI action of the D7 branes \cite{Buchel:2008vz}.
For a realistic value $\lambda_h\sim 6\pi $ both of them are  $\sim 1\%$, although the precise value
will  very much depend on the prefactor, which can vary from one quantity to another. For example, for the free energy, the prefactor is of ${\cal O}(1)$  \cite{Gubser:1998nz},  and therefore we 
expect the previous assertion about the dominance of flavor corrections to hold true for equilibrium quantities, like the entropy. On the other hand,  dynamical quantities, like transport coefficients 
don't necessarily  exhibit this hierarchy. Such is the case, for example, of  the quotient $\eta/s$ which is uncorrected by the flavor, whereas curvature corrections give 
a large prefactor to $\lambda_h^{-3/2}$ of ${\cal O}(10)$  \cite{Buchel:2008ae}.

Flavor corrected thermodynamics shows a departure from conformality  at ${\cal O}(\epsilon_h^2)$, which agrees with expectations, since
we are adding massless flavors and therefore  conformal symmetry is broken by quantum effects.
Being very mild, it would be very interesting to study the hydrodynamics of the plasma, in particular the bulk viscosity. 

We also considered the energy loss of probes in the plasma, by calculating the jet quenching parameter and the friction parameter.
The outcome is that the energy loss is increased by the presence of matter in the fundamental representation.
Even if this statement needs to be supplemented with some prescription of how to compare different theories, we found it to be true in all the reasonable comparison schemes that we analyzed.
Thus, for example, the value of the jet quenching parameter, as compared to the one in ${\cal N}=4$ SYM, is shifted towards the RHIC phenomenological window.
It would be important to understand better the physical reasons behind this result and its implications, if any, for phenomenology.
For example, there are indications that the smaller perturbative cross section of quarks with respect to gluons in the jet quenching process is not consistent with the data, suggesting that the strong coupling cross sections should be quite a bit larger (see for example \cite{Muller:2006ee}, section 3.2).

Besides the straightforward application to the study of phenomena like the ones mentioned above, having analytical control over a perturbative
solution may prove instrumental when trying to obtain an exact backreacted one by numerical methods, so as to provide correct boundary 
and matching conditions.

We finally began to attack the problem of the backreaction of massive quarks.
We derived the equations of motion and briefly considered the zero temperature case.
It is conceivable that a solution at finite temperature can be calculated and analyzed following the methods in this paper.

Apart from the ones outlined above, there are a number of properties and extensions of our solutions that
deserve to be studied.
Basically any observable calculated in ${\cal N}=4$ SYM could be analyzed in the present setting.
Some studies in this direction are currently being performed \cite{noi2}.
Also, one might think about putting the system at finite charge density and studying its properties. 

\vskip 15pt
\centerline{\bf Acknowledgments}
\vskip 10pt
\noindent
We are grateful to N. Armesto,
E. Imeroni,  D. Mateos, C. N\'u\~nez and J. Shock for useful discussions.
This work has been supported by the European Commission FP6 programme
MRTN-CT-2004 v-005104, ``Constituents, fundamental forces and symmetries in the universe''. F. B. is also supported  by the Belgian Fonds de la Recherche Fondamentale Collective (grant 2.4655.07), by the Belgian Institut Interuniversitaire des Sciences Nucl\'eaires (grant 4.4505.86) and
the Interuniversity Attraction Poles Programme (Belgian Science
Policy). A. C. is also supported by the FWO -
Vlaanderen, project G.0235.05 and by the Federal Office for
Scientific, Technical and Cultural Affairs through the Interuniversity Attraction
Poles Programme (Belgian Science Policy) P6/11-P.
The research of A.P is
supported by grants FPA2007-66665C02-02 and DURSI
2009 SGR 168, and by the CPAN CSD2007-00042 project of the
Consolider-Ingenio
2010 program.
A. V. R., J.M. and J.T. are supported by the MEC and  FEDER (grant FPA2008-01838), the Spanish Consolider-Ingenio 2010 Programme CPAN (CSD2007-00042) and Xunta de Galicia (Conselleria de Educacion and grant PGIDIT06PXIB206185PR). J.T. is also supported by MEC of Spain under a grant of the FPU program. 

{ \it F. B. and A. L. C. would like to thank the Italian students, parents and scientists for
their activity in support of public education and research.}

\appendix

\setcounter{equation}{0}
\renewcommand{\theequation}{\Alph{section}.\arabic{equation}}

\section{Technical details regarding the DBI contribution to the effective action}
\label{Qf}
\setcounter{equation}{0}

Let us justify the form of the last term in (\ref{effeclagr}), which comes from the DBI action of the
flavor branes. In the supersymmetric case ($b=1$), we can use the results of \cite{Benini:2006hh}, where  a two-form $\Omega$, related to the distribution density of the flavor branes, was introduced. This form $\Omega$ is  such
that $dF_{(1)}=-g_s \Omega\equiv -g_s \sum_i \Omega^{(i)}$, with the
$\Omega^{(i)}$ being decomposable as the wedge product of two one-forms. Using supersymmetry, it was pointed out in
\cite{Benini:2006hh} that
$S_{DBI} = -T_7 \int d^{10}x \sqrt{-g_{10}}e^\Phi \sum_i \sqrt{\frac12 \Omega_{MN}^{(i)}\Omega_{PQ}^{(i)}
g^{MN}g^{PQ}}$. Since $\Omega$ can be directly computed from the expression of $F_{(1)}$ in (\ref{f5f1}), the $T=0$
version ($b=1$) of the last term in (\ref{effeclagr}) is readily found.

In order to generalize this result to the finite temperature case, one
essentially needs to prove that the massless embeddings of the supersymmetric case are still solutions of
the finite temperature case. With this purpose in mind, let us consider a generic $AdS_5$-BH$\times X_5$ 
background with metric:
\be
ds^2 = h^{-1/2} \left[-b\ dt^2 + d\vec{x}_3\right] + h^{1/2} \left[\frac{dr^2}{b} + r^2 
ds_{KE}^2 + r^2(d\tau + A_{KE})^2\right]\,\,.
\ee
Let us write the K\"ahler-Einstein metric as:
\be
ds_{KE}^2 = d\alpha^2 + \sum_{i=1}^3 w_i(\alpha) \sigma_i^2\,\,,
\ee
where the one forms $\sigma_i$ depend on the other angular coordinates $\xi_a$, $a=1,2,3$. The one-form $A_{KE}$ will depend on $\alpha$ and $\xi_a$ and will not have a component along $d\alpha$.

Let us consider a D7-brane extended along $t, x_i, r, \xi_a$. It is trivial to show that $\tau=const$ is always a solution of the embedding equations, as in the $T=0$ case. Picking $\alpha=\alpha(r)$, the effective DBI lagrangian is proportional to:
\be
{\cal L} \approx r^3 H(\alpha, \xi_a) \sqrt{1+ b(r)r^2 \alpha'^2}\,\,,
\label{effl}
\ee
where $H(\alpha, \xi_a)$ is the square root of the determinant of the metric of the three-dimensional compact space $X_3$ defined as the $\alpha,\tau=const$ slice of $X_5$.

The massless embedding $\alpha=const=\alpha_0$ is a solution of the second order equations following from (\ref{effl}) if:
\be
\partial_{\alpha} H(\alpha, \xi_a) |_{\alpha=\alpha_0} = 0\,\,.
\label{cond}
\ee
This is the same condition one would find in the $T=0$ (i.e. $b=1$) case. Following an analogous reasoning, one can check that $\alpha=\alpha_0$, $\tau=const$
is also a solution of the D7 equations of motion for the backreacted geometries (both $T=0$ and
$T>0$). Using the isometries of the internal manifold, one can generate the family of
embeddings needed for the smearing (both $T=0$ and
$T>0$). Thus, the last term of (\ref{effeclagr}) for $T>0$ is a trivial generalization of the $T=0$ result
quoted above.

Moreover, the relation between $Q_f$ and
$N_f$ can be determined by looking at this last term of (\ref{effeclagr}).
The DBI on-shell action for $N_f$ ``massless'' branes is just:
\be
S_{DBI}= -T_7 N_f \int d^{8}x\, e^{\Phi} \sqrt{-g_8} = - T_7 
N_f Vol(X_3) V_{1,3}\int d\sigma\, e^{\Phi}\, b\, S^6\, F^2\,\,,
\ee
where $Vol(X_3)$ is the volume of the constant $\tau$, constant $\alpha=\alpha_0$ slice of $X_5$
that the D7 wraps, as
discussed
above. In (\ref{effeclagr}), the constant in front of this expression was
defined as $-4Q_f\frac{Vol (X_5)V_{1,3}}{2\kappa_{10}^2}$.
Equating and using $2\kappa_{10}^2 T_7 = g_s$, we find 
the second relation in (\ref{NcNf}).
It is worth noticing that the factor $\frac{Vol(X_5)}{Vol(X_3)}$ appearing in
(\ref{NcNf}) is just the volume transverse to any flavor brane.

\section{Computation of the ADM energy and the free energy}
\label{free_energy}
\setcounter{equation}{0}

Let us derive the result (\ref{ADMresult}).
The ADM energy is given by:\footnote{We use the notations in \cite{Gubser:2001eg}.}
\be
E_{ADM} = -\frac{1}{\kappa_{10}^2}\sqrt{|g_{tt}|} \int d^8 x \sqrt{\det g_{8}} (K_T - K_0)\,\,.
\label{ADM}
\ee
The eight-dimensional integral is taken over a constant time, constant radius hypersurface.
The symbols $K_T$ and $K_0$ are the extrinsic curvatures of the eight-dimensional
subspace within the nine-dimensional (constant time) space, at finite  and
zero temperature, respectively. They are  defined as $K\equiv  \frac{1}{\sqrt{\det g_9}}
\partial_\mu (\sqrt{\det g_9} n^\mu)$ where $n^\mu$ is a normalized vector perpendicular to the surface.
Since we consider constant $r$ hypersurfaces, this is just $n^\mu= \frac{1}{\sqrt{g_{rr}}}\delta^\mu_r$.
The metric on the eight-dimensional
slice must be equal for the zero temperature and finite temperature solutions that are being compared.
With the solution written in (\ref{finiteTsol}), this is the case when $r=r_*$ without any further rescaling of coordinates.
Using (\ref{kappa10}), (\ref{simpleh}), (\ref{deformedads5bh}), the explicit expression for the ADM energy reads:
\be
E=-\frac{\pi N_c^2V_3}{4Vol(X_5) R^8}\frac{r_*}{R}
\left[\frac{r\sqrt{1-\frac{r_h^4}{r^4}}}{R\,\tilde S^4 \tilde F}
\partial_r (r^3 R^2 \tilde S^4 \tilde F)\Big|_{r=r_*}-
\frac{r}{R\,\tilde S_0^4 \tilde F_0}
\partial_r (r^3 R^2 \tilde S_0^4 \tilde F_0)\Big|_{r=r_*}
\right]\,\,.
\ee
Plugging in the solutions (\ref{susymasslesssol}), (\ref{finiteTsol}) we get:
\be
\varepsilon= \frac{E}{V_3}=\frac38 \pi^2 N_c^2 \left(\frac{r_h}{\pi\,R^2}\right)^4
\left[1+\frac{\epsilon_*^2}{24}+O(\epsilon_*^3)\right]\,\,,
\ee
where we have discarded terms suppressed by powers of $\frac{r_h}{r_*}$.
We can now write the result in terms of $\epsilon_h$ and
$T$ by substituting 
(\ref{epsirelation}) and (\ref{temperature}), and obtain (\ref{ADMresult}).

Let us now directly compute the free energy in (\ref{freeen}).
It is given by the Euclidean action ${\cal I}$ 
(renormalized by subtracting the zero temperature result)
evaluated on the solution and divided by the inverse temperature $\beta$ \cite{Hawking:1995fd}. 
We can write:
\be
F=\frac{1}{\beta}({\cal I}_{bulk,T} - {\cal I}_{bulk,T=0} + {\cal I}_{surface,T}
-{\cal I}_{surface,T=0})\,\,.
\ee
We take $r_*$ as the radial cut-off for the integrals, such that the
finite $T$ and zero $T$ geometries coincide. The only subtlety comes from the fact that $g^T_{tt}(r_*)\neq
g^0_{tt}(r_*)$. Thus, in order to compare the solutions, we ought to rescale the Euclidean 
time of the zero
temperature solution such that its period is $\beta_0 = \beta \frac{\sqrt{g^T_{tt}(r_*)}}{\sqrt{g^0_{tt}(r_*)}}
= \beta\sqrt{1-\frac{r_h^4}{r_*^4}}$, where $\beta$ is the period of the Euclidean time of the finite $T$ solution.

The bulk action comes from the sum of (\ref{actiongrav}) and (\ref{actionflav}).
The WZ term in (\ref{actionflav})
does not contribute on-shell for the massless embeddings we are discussing here. By using
the equations of motion of the system, one can prove that the integrand of the on-shell
action is just equal to $-8r^3$. 
We get a very simple expression:
\be
\frac{-2\kappa_{10}^2}{V_3 Vol(X_5)}\left(
{\cal I}_{bulk,T} - {\cal I}_{bulk,T=0}\right)=
\left[\beta \int_{r_h}^{r_*} (-8r^3)dr - \beta \sqrt{1-\frac{r_h^4}{r_*^4}} 
\int_{0}^{r_*} (-8r^3)dr\right] = \beta r_h^4\,\,.
\ee
In this expression and the following, it is understood that corrections in powers
of $\frac{r_h}{r_*}$ and of order $\epsilon_*^3$ or higher are discarded. 
The surface action comes from the Gibbons-Hawking term:
\be
{\cal I}_{surf}= -\frac{1}{\kappa_{10}^2}\int_\Sigma K d\Sigma\,\,,
\ee
where $\Sigma$ is the 9d subspace at $r=r_*$ and $K$ the extrinsic curvature of this hypersurface
within the full 10d space. Explicitly:
\bear
\frac{-2\kappa_{10}^2}{V_3 Vol(X_5)}\left(
{\cal I}_{surf,T} - {\cal I}_{surf,T=0}\right) &=&
2\left[\beta \frac{r\sqrt{1-\frac{r_h^4}{r^4}}}{R\,\tilde S^4 \tilde F}
\partial_r (r^4 R \sqrt{1-\frac{r_h^4}{r^4}}
\tilde S^4 \tilde F)\Big|_{r=r_*}\right.\\
&&\left.-\beta\sqrt{1-\frac{r_h^4}{r_*^4}}
\frac{r}{R\,\tilde S_0^4 \tilde F_0}
\partial_r (r^4 R \tilde S_0^4 \tilde F_0)\Big|_{r=r_*}
\right] = -\frac18 \beta r_h^4 \epsilon_*^2\,\,.\nonumber
\eear
Summing the bulk and surface contributions and inserting
(\ref{kappa10}), (\ref{simpleh}), we get:
\be
F= {\cal I}\,\beta^{-1}= V_3 \frac{\pi^5 N_c^2}{8 Vol(X_5)}\, \left(\frac{r_h}{\pi \,R^2}\right)^4
\left[1-\frac18 \epsilon_*^2 + O(\epsilon_*^3)\right]\,.
\ee
In terms of $\epsilon_h$ and
$T$ from (\ref{epsirelation}), (\ref{temperature}),  we recover (\ref{freeen}).

\section{The smeared Karch-Katz model}
\label{unquenchedkk}
\setcounter{equation}{0}

In this appendix, we want to write down the family of embeddings that participate in the
smearing of massive flavors in $AdS_5\times S^5$ (at the end of the appendix we comment
on the generalization to finite temperature backreacted backgrounds). 
They are all holomorphic embeddings in a given
set of complex coordinates and therefore the overall preserved supersymmetry is ${\cal N}=1$.
From the geometric point of view, rewriting the ${\cal N}=4$ theory in ${\cal N}=1$ components amounts to rewriting the metric of $S^5$ in the canonical Sasaki-Einstein form, i.e. as a $U(1)$ bundle over a complex 4d K\"ahler manifold. In this case, the latter is $CP^2$. 

In the following, we use the notation from \cite{pope}.
Let us consider $\CC^3$ with complex coordinates $Z^1,Z^2,Z^3$ and metric:
\beq
ds^2\,=\,|dZ^1|^2\,+\,|dZ^2|^2\,+\,|dZ^3|^2\,\,.
\label{Z1Z2Z3}
\eeq
We  introduce a radial variable $r$ such that $Z^i = r z^i$, with $r^2 = \sum |Z^i|^2$ and the $z^i$ spanning a unit five-sphere:
\beq
|z^1|^2\,+\,|z^2|^2\,+|z^3|^2\,=\,1\,\,.
\eeq
This constraint is invariant under $SU(3)$ rotations of the $z^i$ and under the transformation
$z^i\to e^{i\alpha}z^i$, where $\alpha$ is a phase.\footnote{Of course, rewriting $Z^i = y^i + i y^{i+3}$ and so on, with $y^j$ real, the defining equation  for $S^5$ reads $\sum_{j=1}^6 {y^j}^2 =1$. This is actually invariant under the larger group $SO(6)\sim SU(4)$. The choice of an holomorphic parameterization breaks this symmetry 
group to the smaller $SU(3)\times U(1)_R$ one.} The space $CP^2$ 
is the space
of orbits under the action of this $U(1)$ group.
We can parameterize:
\bear
z^1=\cos{\chi\over 2}\,\,\cos{\theta\over 2}\,\,
e^{{i\over 2}\,\,(2\tau+\psi+\varphi)}\,\,,\quad
z^2=\cos{\chi\over 2}\,\,\sin{\theta\over 2}\,\,
e^{{i\over 2}\,\,(2\tau+\psi-\varphi)}\,\,,\quad
z^3=\sin{\chi\over 2}\,e^{i\tau}\,\,.
\label{newco}
\eear
By inserting (\ref{newco}) into (\ref{Z1Z2Z3}), one gets the metric of $S^5$ as written
in (\ref{x5metric}), (\ref{cp2metric}).
The gauge theory $U(1)_R$ symmetry is related to the isometry generated by the Killing vector $\partial_\tau$. 
The volume element of $S^5$ in these coordinates is:
\be
\varepsilon(S^5) =
\frac{1}{64}\sin\theta\,\sin\chi(1+\cos\chi)\,d\tau\wedge\,d\psi\wedge
\,d\varphi\wedge\,d\theta\wedge\,d\chi\,\,.
\label{volf}
\ee
The range of the angles is $0\leq \chi, \theta \leq \pi$, 
$0\leq \varphi, \tau < 2\pi$, $0\leq \psi< 4 \pi$
such that 
$\int \varepsilon(S^5) =Vol(S^5)=\pi^3$.
We now  discuss the holomorphic embeddings. We can take
$Z^3 =  r_q $ with $r_q$ real (and related to the modulus of the quark mass)
as the ``fiducial'' (Karch-Katz) embedding \cite{kk}. In terms of the coordinates, the fiducial worldvolume
satisfies
$\, \sin\frac{\chi_{wv}}{2}=\frac{r_q}{r}$ and $\tau=0$.
Notice that $r$ is constrained by $r\ge r_q$ and that it diverges for $\chi=0$.
Acting on it with the $SU(3)\times U(1)_R$ isometry group, we can write the following family of
holomorphic embeddings:
\be
\sum_{i=1}^3 a_i Z^i = r_q \,e^{i\beta}\,\,,\qquad {\rm with}\,\, \sum_{i=1}^3 |a_i|^2=1\,\,,
\label{genem}
\ee
where $0\leq\beta<2\pi$ is associated to the phase of the quark mass and
the $a_i$ are complex parameters. They span a unit five-sphere and
therefore we can represent them by using a parameterization  similar to the one for the 
$z$'s in (\ref{newco}), in terms of the tilded coordinates $\tilde\chi,\tilde\theta,\tilde\tau,\tilde\psi,\tilde\varphi$. Then the generalized Karch-Katz embedding (\ref{genem}) can be rewritten as:
\be
e^{i(\tau+\tilde\tau)}e^{\frac{i}{2}(\psi+\tilde\psi+\varphi+\tilde\varphi)} \Gamma = \frac{r_q}{r} e^{i\beta}\,\,,
\label{explgen}
\ee 
with:
\be
\Gamma \equiv
\cos\frac{\tilde\chi}{2}\cos\frac{\tilde\theta}{2}\cos\frac{\chi}{2}\cos\frac{\theta}{2}+
\cos\frac{\tilde\chi}{2}\sin\frac{\tilde\theta}{2}\cos\frac{\chi}{2}\sin\frac{\theta}{2}\, e^{-i(\varphi+\tilde\varphi)}
+ \sin\frac{\tilde\chi}{2}\sin\frac{\chi}{2}\,e^{-\frac{i}{2}(\psi+\tilde\psi+\varphi+\tilde\varphi)}\,\,.
\ee
The superpotential of the theory reads schematically:
\be
W =  \Phi_1[\Phi_2,\Phi_3] + \tilde q (a_1\Phi_1+a_2\Phi_2+a_3\Phi_3-m) q \,\,,
\ee 
where $\Phi_{1,2,3}$ are the three complex scalars of ${\cal N}=4$ and the $q$'s are the matter multiplets of mass $m\sim r_qe^{i\beta}$ in the fundamental representation provided by the D7-branes.

Following the procedure of \cite{fkw1}, our goal now is to consider a homogeneous, symmetry preserving,
distribution of D7-branes using this six-parameter family of embeddings and compute their charge density
$\Omega$, such that $dF_{(1)}=-g_s \Omega$. Let us start by splitting the complex equation
(\ref{explgen}) into two real ones $f_1=f_2=0$ with:
\be
f_1 = 2(\tau +\tilde\tau) + \psi + \tilde\psi +\varphi + \tilde\varphi + 2 {\rm Arg}[\Gamma] -2\beta + 4\pi n\,\,,\qquad
f_2 = |\Gamma|^2 - \frac{r_q^2}{r^2}\,\,.
\label{f1f2}
\ee
The $\Omega$ is built by summing the 2-forms locally orthonormal to each brane worldvolume \cite{fkw1}:
\be
\Omega=
\int \left[\frac{2 N_f}{(4\pi)^4} \sin\tilde\chi (1+\cos\tilde\chi) \sin\tilde\theta\,
\right]
\Big(\delta(f_1)\delta(f_2) df_1\wedge df_2\Big)\, d\tilde\chi\,d\tilde\theta\,d\tilde\tau\,d\tilde\psi\,d\tilde\varphi\,d\beta\,\,,
\label{omega}
\ee
where the factor in the square brackets corresponds to the
symmetry preserving normalized density of the flavor branes.
In principle, performing the integral (\ref{omega}) would be a challenge but, because of
the preserved 
symmetries, we know that $F_{(1)} = Q_f \, p(r) (d\tau + A_{CP^2})$, as in 
(\ref{massiveforms}), and thus:
\be
\Omega=-\frac{1}{g_s}dF_{(1)}= -\frac{N_f}{2\pi}\left[\,2p(r)\,J_{CP^2}\,+\,
(\d_r p(r)) \,dr\wedge (d\tau + A_{CP^2})  )
\right]\,\,.
\label{omegaap}
\ee
Thus, the problem of solving the integral (\ref{omega}) is reduced to
finding a single function $p(r)$. The easiest way to proceed is to evaluate from
(\ref{omega}) the $dr\wedge d\tau$ component, at $\chi=\pi$. A straightforward
analysis yields (see \cite{fkw1,fkw2,Bigazzi:2008qq} for details of analogous computations in different cases):
\bea
p(r) &=& 0\,\,,\qquad \qquad \qquad \ (r < r_q)\,\,,\rc
p(r) &=& \left(1-\frac{r_q^2}{r^2}\right)^2\,\,,\qquad (r\ge r_q)\,\,.
\label{nfrho}
\eea
As expected, $p(r)$ vanishes at scales below the quark masses while it asymptotes to 1 in the UV.

Up to here, we have written the family of embeddings (\ref{explgen}) and the expression for the corresponding
charge density (\ref{omegaap}), (\ref{nfrho}) when the background is not backreacted and the temperature
vanishes. However, the computation is easily generalized to cases with backreaction,  finite temperature, or
both. 
This is because the isometries  and consequently the integrals over the angles
are exactly the same. One should just change slightly the expression for $f_2$
in (\ref{f1f2})
to be $f_2 = |\Gamma|^2- \sin^2 \chi_{wv}(\sigma)$, where $\chi_{wv}(\sigma)$ parameterizes the fiducial
embedding and, as in the main text, we have used $\sigma$ for the radial coordinate.
The argument that led to (\ref{nfrho}) is readily generalized to obtain (\ref{pofsigma}).


\end{document}